\documentclass[a4paper,11pt]{article}
\pdfoutput=1 % if your are submitting a pdflatex (i.e. if you have
\usepackage{jheppub} % for details on the use of the package, please
\usepackage{amsmath,amssymb,amsthm,amscd,graphicx}
\input epsf.sty

\theoremstyle{definition}

\newcommand{\CI}{{\cal I}}

\newcommand{\CO}{{\cal O}}

\def\IR{{\mathbb R}}
\def\IC{{\mathbb C}}

\def\IP{{\mathbb P}}

\def\IF{{\mathbb F}}

\newcommand{\tr}{{\rm Tr}}
\newcommand{\re}{{\rm e}}
\newcommand{\ri}{\mathsf{i}}
\newcommand{\rd}{{\rm d}}

\newcommand{\mT}{\mathsf{T}}
\newcommand{\mt}{\mathsf{t}}
\newcommand{\mL}{\mathsf{L}}
\newcommand{\mx}{\mathsf{x}}
\newcommand{\my}{\mathsf{y}}
\newcommand{\mm}{\mathsf{p}}
\newcommand{\im}{\mathsf{i}}
\newcommand{\mH}{\mathsf{H}}

\newcommand{\mq}{\mathsf{q}}

\newcommand{\bH}{\boldsymbol{H}}
\newcommand{\fad}{\operatorname{\Phi}_{\mathsf{b}}}
\newcommand{\mb}{{\mathsf{b}}}
\newcommand{\be}{\begin{equation}}
\newcommand{\ee}{\end{equation}}
\newcommand{\ba}{\begin{aligned}}
\newcommand{\ea}{\end{aligned}}
\newcommand{\ben}{\begin{eqnarray}\displaystyle}
\newcommand{\een}{\end{eqnarray}}

\newcommand{\sectiono}[1]{\section{#1}\setcounter{equation}{0}}

\newdimen\tableauside\tableauside=1.0ex
\newdimen\tableaurule\tableaurule=0.4pt
\newdimen\tableaustep
\def\phantomhrule#1{\hbox{\vbox to0pt{\hrule height\tableaurule width#1\vss}}}
\def\phantomvrule#1{\vbox{\hbox to0pt{\vrule width\tableaurule height#1\hss}}}
\def\sqr{\vbox{%
  \phantomhrule\tableaustep
  \hbox{\phantomvrule\tableaustep\kern\tableaustep\phantomvrule\tableaustep}%
  \hbox{\vbox{\phantomhrule\tableauside}\kern-\tableaurule}}}
\def\squares#1{\hbox{\count0=#1\noindent\loop\sqr
  \advance\count0 by-1 \ifnum\count0>0\repeat}}
\def\tableau#1{\vcenter{\offinterlineskip
  \tableaustep=\tableauside\advance\tableaustep by-\tableaurule
  \kern\normallineskip\hbox
    {\kern\normallineskip\vbox
      {\gettableau#1 0 }%
     \kern\normallineskip\kern\tableaurule}%
  \kern\normallineskip\kern\tableaurule}}
\def\gettableau#1{\ifnum#1=0\let\next=\null\else
\squares{#1}\let\next=\gettableau\fi\next}

\newcommand{\figref}[1]{Fig.~\protect\ref{#1}}
\newcommand{\pd}{\partial}
\DeclareMathOperator{\Li}{Li}

\def\({\left(}
\def\){\right)}
\DeclareMathOperator{\real}{Re}

\title{\boldmath Exact quantization conditions for the relativistic Toda lattice}

\author{Yasuyuki Hatsuda}
\author{and Marcos Mari\~no}

\affiliation{D\'epartement de Physique Th\'eorique et Section de Math\'ematiques\\
Universit\'e de Gen\`eve, Gen\`eve, CH-1211 Switzerland}

\emailAdd{Yasuyuki.Hatsuda@unige.ch, Marcos.Marino@unige.ch} 

\abstract{Inspired by recent connections between spectral theory and topological string theory, 
we propose exact quantization conditions for the relativistic Toda lattice of $N$ particles. 
 These conditions involve the Nekrasov--Shatashvili free energy, which resums the perturbative WKB expansion, but they require 
 in addition a non-perturbative contribution, which is related to the perturbative result by an S-duality transformation of the Planck constant. 
 We test the quantization conditions against explicit calculations of the spectrum for $N=3$. Our proposal 
can be generalized to arbitrary toric Calabi--Yau manifolds and might solve the corresponding quantum integrable system of Goncharov and Kenyon.}    

\begin{document}
\maketitle
\flushbottom

\sectiono{Introduction}

An important problem in the theory of quantum integrable systems is to find exact quantization conditions 
determining the eigenvalues of their commuting Hamiltonians. In the case of the 
Toda lattice, these conditions were obtained in \cite{gutzwiller,gp,kl}, by finding appropriate solutions of the one-dimensional quantum Baxter equation
associated to the integrable system. The Toda lattice has a relativistic generalization \cite{ruij}, and the techniques of separation of variables lead to a 
Baxter equation involving difference operators \cite{kuz,kls}. In the relativistic case, the solution to the quantum Baxter equation is not known, even for the 
two-particle lattice. Therefore, the standard techniques of quantum integrable systems have not yet given an explicit solution for the spectrum of the relativistic 
Toda lattice. 

A completely different route to the problem was proposed in \cite{ns}, based on the connection to supersymmetric Yang--Mills theories. As it is well-known, 
the Seiberg--Witten curve of ${\cal N}=2$ Yang--Mills theory in four dimensions is the spectral curve of the Toda lattice \cite{itep,mw}. A similar statement holds for supersymmetric 
Yang--Mills theory in five dimensions, compactified on a circle; in this case, the spectral curve corresponds to the relativistic Toda lattice \cite{nek}. The Bohr--Sommerfeld quantization 
conditions for the corresponding integrable systems are given by periods of the appropriate differential on the spectral curve. These periods can be identified with the dual periods of the 
supersymmetric theory, i.e. with derivatives of the Seiberg--Witten prepotential. 

It was conjectured in \cite{ns} that the all-orders WKB quantization condition of the underlying quantum integrable system can be 
obtained by considering the supersymmetric Yang--Mills theory on the so-called $\Omega$-background, and then taking a special limit, known as the Nekrasov--Shatashvili (NS) limit. 
In this limit, only one out of the two deformation parameters of the $\Omega$ background survives, and it is identified with the Planck constant of the integrable system. The resulting 
one-parameter deformation of the prepotential contains all the quantum corrections to the Bohr--Sommerfeld quantization condition. This has been explicitly verified in examples \cite{mir-mor}. 

In principle, the conjecture of \cite{ns} provides a complete solution to the problem of finding quantization conditions for the integrable systems 
associated to supersymmetric gauge theories. However, one has to be careful, since quantization conditions often require non-perturbative corrections, beyond 
the all-orders WKB solution. A well-known example is the double-well potential in Quantum Mechanics, where the all-orders WKB quantization condition 
requires instanton corrections (see for example \cite{zjj}). A subtler example is the pure quartic potential, in which one has to consider complex instantons \cite{bpv}. 

In the case of the Toda lattice, corresponding to the four-dimensional gauge theory, this problem has an elegant solution, 
since one can {\it resum} the all-orders WKB expansion by using the NS limit of the instanton partition function. 
This partition function can be computed explicitly \cite{nn} as a {\it convergent} power series \cite{ok} in the 
inverse eigenvalues of the commuting Hamiltonians, and each coefficient in this series is exact in $\hbar$. This power series provides then {\it exact} quantization conditions for the Toda lattice. 
Alternatively, one can write these quantization conditions in terms of integral equations of the TBA type \cite{ns}. These TBA 
equations are obtained by resumming the instanton expansion of the partition function (see \cite{my,bourgine} for a detailed derivation), and as shown in \cite{kt} they are 
equivalent to the quantization conditions in \cite{gutzwiller,gp,kl}. 

%In practice, the TBA equations of \cite{ns} are more efficient in order to compute the 

However, in the case of the relativistic Toda lattice, things are more complicated. When $\hbar$ is real, which is the conventional regime for the quantum integrable system, 
the NS limit of the instanton partition function displays an infinitely dense set of poles. As a consequence, the 5d analogue of the 4d quantization condition described above does 
not make sense. The TBA equations of \cite{ns} are only valid, in the 5d case, if ${\rm Im}(\hbar) \not=0$, and they also display poles when $\hbar$ becomes real.
This problem was noted in a similar context in \cite{km} (in fact, the model discussed in \cite{km} can be regarded as a special case of the relativistic Toda lattice with $N=2$ particles). 
Inspired by the HMO mechanism in ABJM theory \cite{hmo}, it was proposed in \cite{km} that these poles should be cancelled 
by non-perturbative contributions in $\hbar$. In the case of 
one-dimensional Hamiltonians associated to quantized mirror curves, the precise form 
of these non-perturbative effects was conjectured in \cite{km, hw, ghm, ghm2,gkmr,wzh,cgm}.

In this paper, based on these works, we propose exact quantization conditions for the relativistic Toda lattice. These conditions involve the 
NS free energy, which is obtained from the NS limit of the instanton partition function and contains the all-orders information about the WKB expansion. 
In addition, they require non-perturbative contributions which cancel the poles of the NS free energy and are related to the 
perturbative all-orders WKB result by an S-duality transformation. As in the cases analyzed in \cite{bpv,km}, these non-perturbative effects have 
their origin in complex instantons appearing in the quantum-mechanical problem. 
The resulting quantization conditions generalize the proposal of \cite{wzh} to the particular 
family of higher genus Calabi--Yau (CY) geometries relevant to the relativistic Toda lattice. As already noted in \cite{hatsuda}, they can be also written 
down for general spectral curves of arbitrary genus. As such, they are 
likely to solve as well the quantum integrable systems associated to arbitrary toric CYs introduced in \cite{gk}. 
 
The organization of this paper is as follows. In section 2 we review some relevant facts about the quantum, relativistic Toda lattice and its realization in terms of CY geometry, and then 
we state our conjectural exact quantization conditions. Section 3 is devoted to a detailed test of these conditions in the case $N=3$. In section 4 we show in detail how the exact quantization 
condition in five dimensions leads to the quantization condition of \cite{ns} in the four-dimensional limit, i.e. for the standard Toda lattice. 
Finally, in section 4 we present some conclusions and prospects for future work.

\sectiono{Quantization conditions for the relativistic Toda lattice}
 
 \subsection{The relativistic Toda lattice}
 
We now review some basic features of the relativistic Toda lattice. Our conventions are similar to those in \cite{kls}, with some 
small modifications. The relativistic Toda lattice describes $N$ interacting particles living on a circle. The position and momentum of the $n$-th particle lead to 
quantum Heisenberg operators $\mq_n,\,\mm_n$ satisfying the
standard commutation relations: 
\be
[\mq_n,\mm_m]=\ri \hbar_{\rm RT} \delta_{nm}, \qquad  n, m=1, \cdots, N. 
\ee
The model depends on a real parameter $R$, and we will denote 
\be
\label{hq}
\hbar= R \hbar_{\rm RT},\qquad  q=\re^{\im \hbar}. 
\ee
 The quantum theory is defined by the Hamiltonian
 \be\label{trm4a'}
\mH(\mq_1,\mm_1;\ldots;\mq_{N},\mm_{N})=\sum_{n=1}^N
\Big\{1+q^{-1/2} R^2 \re^{\mq_n-\mq_{n+1}}\Big\}\,\re^{R \mm_n}, 
\ee
where the variables are periodically identified: $\mq_{N+1}=\mq_1$. It turns out that this model is integrable, and it has $N-1$ commuting Hamiltonians. 
Their explicit expression can be obtained by considering the Lax operator 
\be
\label{laxz}
\mL_n (z;R) = \begin{pmatrix} z- z^{-1} \re^{R \mm_n} & R \re^{-\mq_n} \\
- R \re^{\mq_n + R \mm_n} & 0\end{pmatrix}. 
\ee
The corresponding monodromy matrix is 
\be
\mT(z; R)=\mL_N (z;R) \cdots \mL_1(z;R), 
\ee
and it can be shown that
\be
2 \mt(z;R) = \tr \, \mT(z;R)
\ee
satisfies the commutation relation 
\be
[\mt(z;R), \mt(w;R)]=0. 
\ee
This means that the coefficients $ \mH_k$ in the polynomial 
\be
2 \mt(z;R)=  \sum_{k=0}^N (-1)^k z^{N-2 k} \mH_k \left(\mq_1,\mm_1;\ldots;\mq_{N},\mm_{N} \right)
\ee
mutually commute. Note that $\mH_0=1$ and $\mH_1= \mH$ is the Hamiltonian. We also have that
\be
\mH_{N-1}=  \sum_{n=1}^N
\left\{1+q^{-1/2} R^2 \re^{\mq_{n-1}-\mq_{n}}\right\}\,\re^{-R \mm_n}, 
\ee
and
\be
\mH_{N}= \exp \left( \sum_{n=1}^N \mm_n \right). 
\ee
We can mod out the motion of the center of mass by fixing the total momentum to be zero, so that $\mH_{N}=1$. In this way we have $N-1$ non-trivial commuting 
Hamiltonians $\mH_1, \cdots, \mH_{N-1}$. 

The conventional Toda lattice is recovered in the limit $R \rightarrow 0$. One finds, 
\be
\ba
\mH(\mq_1,\mm_1;\ldots;\mq_{N},\mm_{N}) &= N + R \sum_{n=1}^N\mm_n+ R^2  \sum_{n=1}^N \left( {\mm_n^2\over 2} + \re^{\mq_n-\mq_{n+1}} \right)
\\ & +R^3 \sum_{n=1}^N \left({\mm_n^3 \over 6} + \re^{\mq_n - \mq_{n+1}} \left( \mm_n  - {\im \hbar \over 2} \right) \right)+ \CO(R^4), \\
\mH_{N-1}(\mq_1,\mm_1;\ldots;\mq_{N},\mm_{N}) &= N - R \sum_{n=1}^N\mm_n+ R^2  \sum_{n=1}^N \left( {\mm_n^2\over 2} + \re^{\mq_n-\mq_{n+1}} \right)
\\ & -R^3 \sum_{n=1}^N \left({\mm_n^3 \over 6} + \re^{\mq_{n-1} - \mq_{n}} \left( \mm_n  + {\im \hbar \over 2} \right) \right)+ \CO(R^4). 
\ea
\ee
At quadratic order in $R$ we find the standard Hamiltonian of the Toda lattice, and higher order terms in $R$ lead to the higher Hamiltonians. Equivalently, we can 
set
\be
\label{z-mu}
z= \re^{R \mu/2}, 
\ee
and consider the $R\rightarrow 0$ limit of the Lax matrix (\ref{laxz}), 
\be
\mL_n(z;R) \approx R\begin{pmatrix} \mu- \mm_n & \re^{-\mq_n} \\
-\re^{\mq_n} & 0 \end{pmatrix}= R \mL_n (\mu), 
\ee
 where $\mL_n(\mu)$ is exactly the Lax operator of the non-relativistic Toda lattice (see for example \cite{kl}). 
 %In particular, we find that 
 %
% \be
% 2 t(z; R) \approx R^N 2 t(\mu), \qquad  R\rightarrow 0. 
% \ee
 %
 
 In order to formulate the spectral problem for the quantum relativistic Toda lattice, we eliminate the motion of the center of mass in favor of $N-1$ coordinates $\zeta_1, \cdots, \zeta_{N-1}$. 
Solving the spectral problem means then finding square integrable functions $\psi_{\bH} (\zeta_1, \cdots, \zeta_{N-1}) \in L^2(\IR^{N-1})$, 
 labelled by an $(N-1)$-tuple of eigenvalues
 \be
 \bH= (H_1, \cdots, H_{N-1}),
 \ee
 and such that
 \be
 \label{eigen-prob}
 \mt (z;R)\psi_{\bH}(\zeta_1, \cdots, \zeta_{N-1})= t(z;R) \psi_{\bH}(\zeta_1, \cdots, \zeta_{N-1}), 
 \ee
 where 
 \be
 2 t(z;R)= \sum_{k=0}^N (-1)^k z^{N-2 k} H_k. 
 \ee
 It is convenient to parametrize this polynomial as 
 \be
 2 t(z;R)= \prod_{i=1}^{N}  2 \sinh \left( R {\mu- \mu_i \over 2}\right). 
\ee
The spectral curve for the relativistic Toda lattice is then given by 
\be
\label{rt-sc}
R^N\left( \re^\xi + \re^{-\xi} \right)+ 2 t(z;R)=0, 
\ee
and in the limit $R \rightarrow 0$ we recover the spectral curve of the standard Toda lattice,  
\be
\re^\xi + \re^{-\xi}  + \prod_{i=1}^{N} \left( \mu -\mu_i \right) =0. 
\ee

The method of separation of variables, when applied to the relativistic Toda lattice, implies that the eigenvalue problem (\ref{eigen-prob}) 
can be solved by considering the ``quantum" version of the spectral curve \cite{sklyanin,kuz, kls}. 
We impose the commutation relation
\be
[\mu, \xi]=\ri \hbar_{\rm RT}.
\ee
Then the variable $\xi$ is expressed by a differential operator, 
\be
\label{xi-dif}
 \xi\rightarrow  -\im \hbar_{\rm RT} {\rd \over \rd \mu}, 
\ee
so that its exponential becomes a functional difference operator acting on a ``wavefunction" $\psi(\mu)$. The resulting one-dimensional, quantum Baxter equation reads
\be
\label{q-baxter}
R^N \left( \psi(\mu+ \im \hbar_{\rm RT})+ \psi(\mu- \im \hbar_{\rm RT}) \right) + 2t\left(\re^{R\mu/2}; R \right)\psi(\mu)=0. 
\ee
In principle, the eigenvalue problem (\ref{eigen-prob}) can be solved by solving this equation. However, to do this, one has to specify very carefully the boundary 
conditions satisfied by $\psi(\mu)$, as it happens for example 
in the simpler case of the standard Toda lattice. In this paper, instead of solving this equation analytically, we will propose an exact quantization condition
for the eigenvalues $H_1, \cdots, H_{N-1}$, based on insights from \cite{ns} and on the recent progress in the quantization of mirror curves \cite{km,hw,ghm}. 

\subsection{The Calabi--Yau geometry}

The spectral curve of the relativistic Toda lattice (\ref{rt-sc}) 
can be regarded as the mirror curve for a special CY geometry $X_{N-1}$ which is an $A_{N-1}$ fibration over $\IP^1$. 
There is a well-known connection between topological string theory on this CY geometry, and the relativistic Toda lattice. This connection goes as follows. 
According to \cite{ns}, the perturbative quantization condition for the relativistic Toda lattice with $N$ particles 
is encoded in the instanton partition function for the 5d supersymmetric Yang--Mills theory with gauge group $SU(N)$. In addition, this theory can be engineered by using 
topological string theory on the $X_{N-1}$ geometry \cite{kkv}. In fact, the instanton partition function obtained in \cite{nn} agrees exactly with the 
topological string partition function \cite{ikp, ikp2,taki}, as computed by the (refined) topological vertex \cite{akmv,ikv}. Therefore, we conclude that the WKB quantization 
condition can be obtained from the refined topological string partition function on the $X_{N-1}$ geometry, in the NS limit. 

As we will see, in order to find an exact quantization condition for the relativistic Toda lattice, it is very convenient to use the topological string 
theory point of view. We will now review some basic, general results on (refined) topological strings, and some more concrete results on the $X_{N-1}$ geometry which 
is relevant for our study. 

The topological string free energy on a CY manifold $X$, when expanded around the large radius point, 
is the sum of two pieces. The first, ``perturbative" piece, involves a cubic polynomial in the K\"ahler moduli $T_i$, $i=1, \cdots, n$, plus the so-called constant 
map contribution. The second piece can be regarded as a generating functional of BPS invariants of the CY \cite{gv2,ikv}. These invariants 
arise naturally when one considers M-theory compactified on $X$. In this compactification, M2 branes 
wrapping a two-cycle of $X$ with degree ${\bf d}$ lead to BPS states in five dimensions, 
with spins $(j_L, j_R)$ with respect to the rotation group $SU(2)_L \times SU(2)_R$. The index for such states, 
which we denote by $N^{\bf d}_{j_L, j_R}$, is a topological invariant in the case of local CY manifolds. To write such an index, let us introduce the $SU(2)$ character, 
\be
\chi_{j}(q)
={q^{2j+1}-q^{-2j-1} \over q-q^{-1}}. 
\ee
Notice that, with our conventions, $j_L$ and $j_R$ are generically half-integers.  
The refined topological string free energy is a function of the K\"ahler moduli and of two 
parameters $\epsilon_{1,2}$, which ``refine" the topological string coupling constant. We also introduce (see for example \cite{hk,ckk}) 
\be
\epsilon_L={\epsilon_1 -\epsilon_2 \over 2},\qquad
\epsilon_R= {\epsilon_1 +\epsilon_2 \over 2},
\ee
and
\be
\label{qt}
q=\re^{ \im \epsilon_1},   \qquad t=\re^{-\im \epsilon_2}, \qquad q_{L,R}= \re^{\im \epsilon_{L,R}}.
\ee
(Since we will eventually identify $\epsilon_1=\hbar$, the variable $q$ appearing here is precisely the same one introduced in (\ref{hq})). 
Then, the BPS part of the refined topological string free energy can be written as
\be
\label{refBPS}
F^{\rm BPS}({\bf Q}, \epsilon_1, \epsilon_2)=
-\sum_{j_L, j_R \ge 0 }\sum_{w\ge 1}\sum_{{\bf d}} 
{1\over w}  N_{j_L, j_R}^{\bf d}  {\chi_{j_L} (q^w_L) \chi_{j_R} (q^w_R) \over \left(q^{w/2} - q^{-w/2}\right) \left(t^{w/2} -t^{-w/2} \right)}
{\bf Q}^{w{\bf d}}. 
\ee
In this formula, we have denoted
\be
Q_i=\re^{-T_i}, \qquad i=1, \cdots, n,  \qquad {\bf Q}^{{\bf d}}=\prod_{i=1}^n Q_i^{d_i}. 
\ee
It is also very useful to introduce another set of integer invariants $n_{g_L, g_R}^{\bf d}$ by the following equality of generating functionals, 
\be
\label{change-basis}
\ba
&\sum_{j_L, j_R\ge 0}  N_{j_L, j_R}^{\bf d}
\chi_{j_L}(q_L) \chi_{j_R}(q_R)\\
&=\sum_{g_L, g_R \ge 0 }  n_{g_L, g_R}^{\bf d}
\left(q_L^{1 /2} - q_L^{-1 /2} \right)^{2g_L}
\left(q_R^{1 /2} - q_R^{-1 /2} \right)^{2g_R}.
\ea
\ee
In terms of these invariants, the generating functional (\ref{refBPS}) reads, 
\be
\label{ref-free-2}
F^{\rm BPS}({\bf Q}, \epsilon_1, \epsilon_2)=
-\sum_{g_L, g_R \ge 0 }\sum_{w\ge 1}\sum_{{\bf d}} 
{1\over w} n_{g_L, g_R}^{\bf d}  
{ \left(q_L^{w /2} - q_L^{-w /2} \right)^{2g_L} \over q^{w/2} - q^{-w/2} }
{ \left(q_R^{w /2} - q_R^{-w /2} \right)^{2g_R} \over t^{w/2} -t^{-w/2} }
{\bf Q}^{w{\bf d}}.
\ee
%
%Sometimes it is also useful to consider the perturbative expansion of the free energy, which following \cite{hk} we will write as
%
%\be
%\label{ref-pert}
% F({\bf Q}, \epsilon_1, \epsilon_2)= \sum_{n,g\ge 0} \left( \epsilon_1+ \epsilon_2 \right)^{2n} \left( \epsilon_1 \epsilon_2\right)^{g-1} F^{(n,g)}( {\bf Q}).
% \ee
 %
 
The standard topological string is a particular case of the refined topological string, corresponding to 
\be
\epsilon_1=-\epsilon_2=g_s. 
\ee
In this limit, $q_R=1$, and the only invariants which survive in (\ref{ref-free-2}) have $g_R=0$ and coincide with the Gopakumar--Vafa invariants \cite{gv2}, which we will denote by 
\be
n_g^{\bf d}= n_{g,0}^{\bf d}.
\ee
The total free energy 
of the standard topological string has a genus expansion 
\be
F({\bf Q}, g_s)= \sum_{g\ge 0} F_g({\bf Q}) g_s^{2g-2}. 
\ee
The leading term as $g_s \rightarrow 0$, $F_0({\bf Q})$, is called the genus zero free energy or {\it prepotential} of the CY $X$. It is given by 
\be
F_0({\bf Q})=F^{\rm pert}_0({\bf Q})+ F^{\rm BPS}_0({\bf Q}), 
\ee
where the first term in the r.h.s is the perturbative part, 
\be
\label{per-pre}
F^{\rm pert}_0({\bf Q})={1\over 6} \sum_{i,j,k=1}^n a_{ijk} T_i T_j T_k,
\ee
 while the second one is a sum over worldsheet instantons and involves the genus zero Gopakumar--Vafa invariants: 
\be
F_0^{\rm BPS}({\bf Q})= \sum_{{\bf d}}\sum_{w\ge1} {n_0^{\bf d} \over w^3} {\bf Q}^{w {\bf d}}. 
\ee

There is another special limit of the refined topological string, which was first identified in \cite{ns} and will be the relevant one for our purposes. In this limit, 
one of the epsilon parameters goes to zero while the other is kept finite,
\be
 \epsilon_1=\hbar, \qquad \epsilon_2 \to 0.
\ee
This is usually called the NS limit. The refined free energy (\ref{ref-free-2}) has a simple pole in this limit, and in order to extract a finite piece 
we consider
\be
\ba
\label{lim-ns}
F^{\rm NS, \, BPS}({\bf Q}, \hbar) &=- \lim_{\epsilon_2 \rightarrow 0} 
\epsilon_2 F^{\rm BPS} ({\bf Q}, \epsilon_1, \epsilon_2)\\
&= \sum_{j_L, j_R} \sum_{w, {\bf d} } 
N^{{\bf d}}_{j_L, j_R}  \frac{\sin\frac{\hbar w}{2}(2j_L+1)\sin\frac{\hbar w}{2}(2j_R+1)}{2 w^2 \sin^3\frac{\hbar w}{2}} {\bf Q}^{w {\bf d}}, 
\ea
\ee
which we expressed in terms of the BPS invariants appearing in (\ref{refBPS}). We now define the NS free energy as the sum 
\be
F^{\rm NS}({\bf Q}, \hbar) = F^{\rm NS,\, pert }({\bf Q}, \hbar)+ F^{\rm NS, \, BPS}({\bf Q}, \hbar),
\ee
where the first summand in the r.h.s. is the perturbative part, 
\be
 F^{\rm NS,\, pert }({\bf Q}, \hbar)= {1\over 6 \hbar} \sum_{i,j,k=1}^n a_{ijk} T_i T_j T_k + \sum_{i=1}^n b_i^{\rm NS}\left( { 4 \pi^2 \over \hbar} + \hbar  \right) T_i . 
 \ee
 The leading order term agrees with (\ref{per-pre}). The coefficients $b_i^{\rm NS}$ 
 can be determined by using for example the refined holomorphic anomaly equations of \cite{hk,kw,hkk}. The BPS part of the NS free energy (\ref{lim-ns}) can be also 
computed in terms of another set of integer invariants, 
\be
\label{NS-GV}
F^{\rm NS, \, BPS}({\bf Q}, \hbar)=\ri \sum_{g=0}^\infty \sum_{w\ge 1} \sum_{{\bf d}} 
{1\over w^2}\hat n_{g}^{\bf d}  {  \left(q ^{w /4} - q^{-w /4} \right)^{2g } 
\over q^{w/2} -q^{-w/2} } {\bf Q}^{w{\bf d}},
\ee
where
\be
\label{ns-invs}
\hat n_g^{{\bf d}}= \sum_{g_L+g_R=g} n_{g_L, g_R}^{{\bf d}}.
\ee
By expanding the NS free energy around $\hbar=0$, one finds the following power series, 
\be
\label{hbar-ex}
F^{\rm NS}({\bf Q}, \hbar)=\sum_{n \ge 0} \hbar^{2n-1} F^{\rm NS}_n ({\bf Q}). 
\ee
Note that the first term in this expansion is equal to the prepotential of the CY manifold $X$, up to a linear term in $T_i$: 
\be
F^{\rm NS}_0({\bf Q})= F_0({\bf Q})+ 4 \pi^2 \sum_{i=1}^n b_i^{\rm NS} T_i.
\ee

Let us now consider in more detail the toric CY geometry $X_{N-1}$. It can be described as a resolution of the cone over the Sasaki--Einstein manifold $Y^{N,0}$ (see for example 
\cite{bt}), or as the $A_{N-1}$ fibration over $\IP^1$ with Chern--Simons invariant $m=0$ (see for example \cite{ikp2,ks}). In the standard toric descriptions, it is given by the set of 
$N$ charge vectors in $\IC^{N+3}$, 
\be
\ba
e_1&=(0,0, 1,-2,1,0, 0,\cdots,0,0,0 ,0), \\
e_2&=(0,0,0,1,-2,1, 0,\cdots, 0,0,0,0),\\
\vdots & \qquad \qquad \qquad \vdots \\
e_{N-1}&= (0,0,0, 0, 0,0,0,\cdots, 0,1,-2,1),\\
e_N&=(1,1,-1,0,0,0,0,\cdots, 0,0,0,-1).
\ea
\ee
\begin{center}
 \begin{figure}
\begin{center}
 {\includegraphics[scale=0.55]{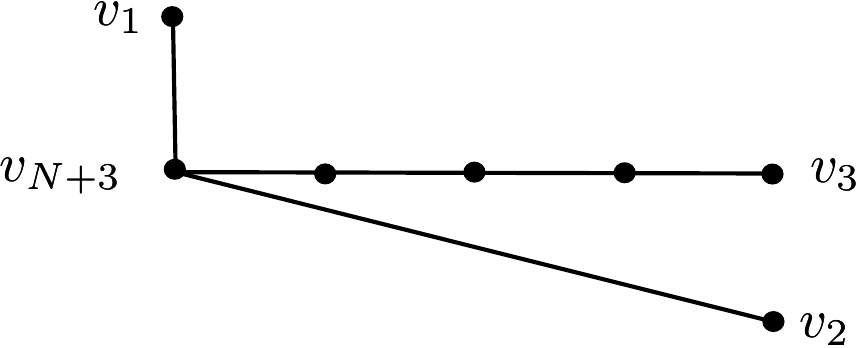}}
\caption{A height one slice of the toric fan (\ref{t-fan}).}
 \label{fan}
\end{center}
\end{figure}  
\end{center}
The corresponding fan is
\be
\label{t-fan}
\ba
v_1&=(0,1,1), \\
v_2&=(N, -1, 1), \\
v_{i+3}&=(N-i, 0, 1), \qquad i=0, \cdots, N. 
\ea
\ee
A graphic representation of this fan in the case $N=4$ can be seen in \figref{fan}. This determines the mirror curve $\Sigma$ to be 
\be
\label{genmc}
b_1\re^{p}+ b_2 \re^{Nx-p}+\sum_{i=0}^N b_{i+3} \re^{(N-i)x}=0. 
\ee
This curve has genus $g_\Sigma=N-1$. The corresponding Batyrev coordinates for the moduli space are 
\be
\ba
z_i&={b_{i+2} b_{i+4} \over b_{i+3}^2}, \qquad i=1, \cdots, N-1, \\
z_N&={b_1 b_2 \over b_3 b_{N+3}}. 
\ea
\ee
It turns out that $z_1, \cdots, z_{N-1}$ are true moduli of the mirror curve, while $z_N$ is a mass parameter, 
in the terminology of \cite{hkp,hkrs}. Mirror symmetry determines the K\"ahler parameters $T_i$ to be given by the mirror map, 
\be
\label{mir-map}
\ba
-T_i&=\log(z_i) + \CO(z_i), \qquad i=1, \cdots, N-1, \\
-T_N&=\log(z_N). 
\ea
\ee
Note that the mirror map for the mass parameter is algebraic. 
The perturbative part of the NS free energy can be computed by various methods, e.g. the 5d instanton partition function in \cite{no}, 
and one finds
\be
\label{pertNS}
\ba
\hbar F^{\rm NS}({\boldsymbol{T}}, \hbar) &={1\over 6} \sum_{1\le l < n\le N-1} (T_l +T_{l+1}+ \cdots + T_n)^3 +{T_N\over 2 N}  \sum_{1\le l <n\le N-1} (T_l +T_{l+1} +\cdots + T_n)^2\\
&- \left( {\pi^2 \over 3} 
+ {\hbar^2 \over 12} \right) \sum_{l=1}^{N-1} l(N-l) T_l+\cdots,
\ea
\ee
where the $\dots$ indicate the BPS generating function (\ref{lim-ns}). This function can be also derived by many different methods \cite{hk,kw,hkk,ckk}, but for our purposes 
it is more convenient to use the (refined) topological vertex \cite{ikp,ikp2,taki}, or, equivalently, Nekrasov's five dimensional instanton 
partition function \cite{nn,no}. We will spell out 
some details of such a calculation in the case $N=3$, in the next section. Note that 
\be
b_i^{\rm NS}= -{i(N-i) \over 12}, \qquad i=1, \cdots, N-1. 
\ee

Another important ingredient in our calculation is the quantum mirror map, which was introduced in \cite{acdkv}. This 
promotes the non-trivial K\"ahler parameters $T_i$, $i=1, \cdots, N-1$, to functions of 
the $z_i$, $i=1,\cdots, N$, and of $\hbar$. We will denote the resulting functions by $T_i(z_1, \cdots, z_N; \hbar)$ or simply by $T_i(\hbar)$. 
As explained in \cite{acdkv}, the quantum mirror map can be computed as follows. Let us denote the equation for the mirror curve (\ref{genmc}) by $H(x, p)=0$. Upon quantization, the momentum 
$p$ becomes a differential operator, 
\be
p \rightarrow -\im \hbar {\rd \over \rd x} 
\ee
and the equation $H(x,p)=0$ becomes a functional difference equation
\be
\label{qc-gen}
H \left( x, -\im \hbar {\rd \over \rd x}  \right) \psi(x)=0,
\ee
which can be solved for $\psi(x)$ as a power series in the moduli. The quantum mirror map is then obtained by computing periods of $\log \psi(x)$. 
In section \ref{testing} we will perform a detailed calculation of the quantum mirror map in the case $N=3$. 

%This leads to the functional difference equation, 
%
%\be
%b_1 V(X)+ { b_2 q^{N/2} X^N  \over V(q X)}+ \sum_{i=0}^N b_{i+3}X^{N-i}=0, 
%\ee
%
%where
%
%\be
%V(X)= {\psi\left(q^{-1} X \right) \over \psi(X)}. 
%\ee
%

 \subsection{Exact quantization conditions}

\begin{center}
 \begin{figure}
\begin{center}
 {\includegraphics[scale=0.6]{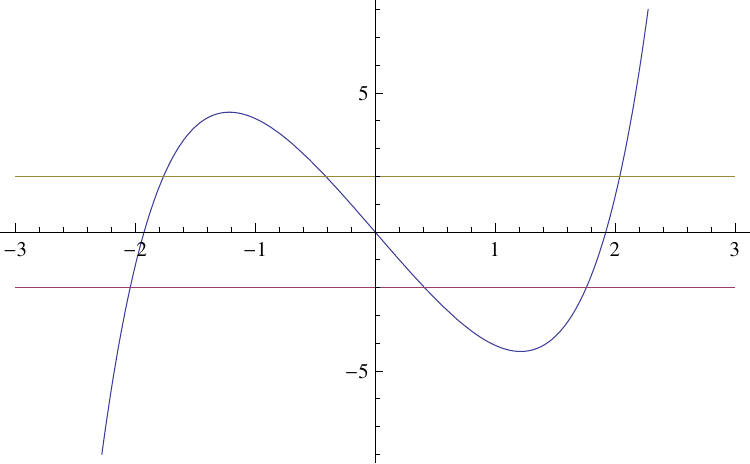}}
\caption{The intervals of instability for the relativistic Toda lattice with $N=3$ particles, for $H_1=H_2=8$ and $R=1$.
In this case, there exist two finite intervals ${\cal I}_1$ and ${\cal I}_2$.}
 \label{R-isfig}
\end{center}
\end{figure}  
\end{center}

A first approach to the quantization conditions in an integrable system is to use the WKB method, i.e. the Bohr--Sommerfeld approximation and its 
all-orders generalization (in the context of higher dimensional integrable system, the Bohr--Sommerfeld approximation is sometimes known as the 
Einstein--Brillouin--Keller, or EBK, method.) This approximation
 is based on the action variables for the classical motion. 
By analogy with the standard Toda lattice (see for example the excellent presentation in \cite{bbt}), we define the 
intervals of instability as those intervals $\CI_k \subset \IR$, $k=1, \cdots, N-1$, where 
\be
{\left| t(z;R) \right| \over R^N}\ge 1.
\ee
An example for the $N=3$ relativistic Toda lattice can be found in \figref{R-isfig}. 
The action variables are then given by 
\be
\label{rt-actions}
I_k=2  \int_{\CI_k} \cosh^{-1} \left|  {t(z;R) \over R^N}  \right| \rd \mu, \qquad k=1, \cdots, N-1, 
\ee
and the Bohr--Sommerfeld quantization condition reads 
\be
\label{bs-qc}
{1\over 2 \pi \hbar_{\rm RT}} I_k = n_k+{1\over 2}, \qquad k=1, \cdots, N-1. 
\ee
The action integrals are B-periods of the appropriate meromorphic differential on the mirror curve, and there should be given by linear combinations 
of the periods of the CY geometry. We find that, 
\be
R I_k = \sum_{j=1}^{N-1} C_{kj} \left( {\partial \widehat F_0 \over \partial T_j}+4 \pi^2 b^{\rm NS}_j\right). 
\ee
In this equation, $C_{kj}$ is the Cartan matrix of $SU(N)$, and the hatted prepotential is defined as
\be
\label{hatF}
 \widehat F_0({\bf Q})= F_0^{\rm pert}({\bf Q})+ F_0^{\rm BPS}( Q_1, \cdots, Q_{N-1}, (-1)^N Q_N). 
 \ee
 The K\"ahler parameters $T_i$ are related to the moduli $z_i$ by the mirror map (\ref{mir-map}), while the moduli are related to the parameter $R$ and Hamiltonians by 
\be
\label{z-H}
\ba
z_i&={H_{i-1} H_{i+1} \over H_i^2}, \qquad i=1, \cdots, N-1, \\
z_N&=R^{2N} . 
\ea
\ee
Remember that $H_0=H_N=1$. The relationship (\ref{z-H}) can be obtained by 
comparing the spectral curve (\ref{rt-sc}) with the mirror curve (\ref{genmc}). This comparison also shows that there should be a sign $(-1)^N$ in $Q_N=z_N$ in (\ref{hatF}). 

The Bohr--Sommerfeld approximation can be improved by considering the all-orders WKB result. 
In Appendix~\ref{sec:WKB}, we explain how to compute the perturbative quantum corrections to the classical periods
by using Baxter's difference equation \eqref{q-baxter}.
According to \cite{ns}, these corrections 
are captured by the expansion of the NS free energy around $\hbar=0$. We then find an all-orders WKB quantization condition, 
\be
\label{all-WKB}
\sum_{j=1}^{N-1} C_{kj} {\partial  \over \partial T_j} \widehat F^{\rm NS}\left (\widehat{\boldsymbol{T}}(\hbar), \hbar \right) = 2 \pi  \left( n_k+{1\over 2} \right), \qquad k=1, \cdots, N-1, 
\ee
where $n_k=0,1, 2, \cdots$ are non-negative integers. Here, $\widehat F^{\rm NS}({\bf Q}, \hbar)$ is defined by an equation similar to (\ref{hatF}), 
\be
\label{hatfns}
 \widehat F^{\rm NS} ({\bf Q}, \hbar)= F^{\rm NS, \, pert} ({\bf Q}, \hbar)+ F^{\rm NS, \, BPS} ( Q_1, \cdots, Q_{N-1}, (-1)^NQ_N,\hbar). 
 \ee
Note that the $\hbar$ parameter appearing in the NS free energy is identified with $R \hbar_{\rm RT}$, as in (\ref{hq}). 
The K\"ahler parameters $T_i$, $i=1, \cdots, N-1$ are related to the moduli $z_1, \cdots, z_N$ (therefore to the radius $R$ and the Hamiltonians) 
through the {\it quantum} mirror map $T_i(\hbar)$, but 
after changing $z_N \rightarrow (-1)^Nz_N$, as we noted above. The resulting quantum mirror map has been denoted by 
\be
\ba 
\widehat T_i(z_1, \cdots, z_{N-1}, z_N; \hbar)&=T_i (z_1, \cdots, z_{N-1}, (-1)^N z_N; \hbar), \qquad i=1, \cdots, N-1, \\
\widehat T_N&=T_N = \log z_N. 
\ea
\ee

Equation (\ref{all-WKB}) is essentially the quantization condition stated in \cite{ns}. However, in order to turn it into an exact statement, one has to be 
more precise about the structure of $F^{\rm NS}$. If we consider the NS free energy to be given by its expansion (\ref{hbar-ex}) in powers of $\hbar$, then the l.h.s. is an asymptotic 
series. In this case, unless further information is given, (\ref{all-WKB}) is only an {\it approximate} quantization condition. One can try to improve the situation by considering a 
trans-series encoding non-perturbative information, and by giving a Borel resummation 
prescription for the full trans-series. We will not explore this route in this paper. 

Another possibility is to use the BPS structure of the NS free energy (\ref{lim-ns}). This effectively resums the $\hbar$ expansion and provides a 
series in powers of the $Q_i$, or equivalently in inverse powers of the eigenvalues $H_i$, $i=1, \cdots, N-1$. Unfortunately, the coefficients of the series (\ref{lim-ns}) have 
a dense set of poles on the real $\hbar$ axis. These occur at values of the form 
 \be
 \label{rat-hbar}
 \hbar= 2 \pi {r \over s}, 
 \ee
where $r,s$ are coprime integers\footnote{The 5d instanton partition function of \cite{nn} provides a resummation of the series (\ref{lim-ns}) in the ``fiber" parameters $Q_i$, $i=1, \cdots, N-1$, but 
this does not remove the poles at (\ref{rat-hbar}), which are independent of the K\"ahler moduli.}. 
This fact was first pointed out in a closely related context in \cite{km}, based on insights from \cite{hmo}. 
These poles are not physical, and they should be cancelled by non-perturbative contributions. In order to find 
these contributions, we will use recent results on the quantization of mirror curves 
from \cite{km,hw,ghm}. It turns out that the quantum Baxter equation (\ref{q-baxter}) in the case $N=2$ is identical to the quantization of the mirror curve of local 
$\IP^1 \times \IP^1$. A conjectural, analytic solution to the corresponding eigenvalue problem was presented in \cite{km} for the 
so-called ``maximally supersymmetric case" $\hbar=2\pi$, and then 
extended to arbitrary $\hbar$ in \cite{ghm}. This solution involves the resummation of the all-orders WKB result, but it incorporates in addition 
explicit non-perturbative contributions that cancel the poles. In \cite{wzh} an alternative, conjecturally equivalent form for this quantization condition was found. 
As pointed out in \cite{hatsuda}, the formulation of \cite{wzh} 
has a natural extension to the higher genus case, which corresponds to $N \ge 3$ in the relativistic Toda lattice. In the formulation of \cite{wzh}, the non-perturbative 
contribution is given by an S-duality transformation of the all-orders WKB result. 

In order to see how this works, and to open the way for generalizations, let us consider a general, toric CY manifold $X$ and 
let us introduce, as in \cite{hmmo,ghm,cgm}, a B-field $\boldsymbol{B}$ satisfying the following requirement: 
for all ${\bf d}$, $j_L$ and $j_R$ such that the BPS invariant $N^{{\bf d}}_{j_L, j_R} $ is non-vanishing, we must have
\be
\label{B-prop}
(-1)^{2j_L + 2 j_R+1}= (-1)^{\boldsymbol{B} \cdot {\bf d}}. 
\ee
For local del Pezzo CY threefolds, the existence of such a vector was established in \cite{hmmo}. Let us now consider the combination 
\be
\label{combi}
{\partial \over \partial T_k} F^{\rm NS, BPS}\left(\boldsymbol{T}+ \pi \ri \boldsymbol{B}, \hbar \right)+
\frac{\hbar}{2\pi}
 {\partial \over \partial T_k} F^{\rm NS, BPS}\left({2 \pi \over \hbar} \boldsymbol{T}+ \pi \ri \boldsymbol{B}, {4 \pi^2  \over \hbar} \right), \qquad k=1, \cdots, n. 
\ee
Let us call the first term the WKB contribution, and the second term the non-perturbative contribution. 
The WKB contribution is a power series in the $Q_i$, $i=1, \cdots, n$, and its coefficients have poles when $\hbar$ is of the form (\ref{rat-hbar}). The non-perturbative 
contribution is a power series in the $Q_i^{2 \pi/\hbar}$, $i=1, \cdots, n$, and its coefficients have poles at the same set of values of $\hbar$. 
We will now verify that, if the B-field satisfies (\ref{B-prop}), the poles in the WKB contribution cancel against the poles in the 
non-perturbative contribution. Indeed, the poles at $\hbar =2 \pi r/s$ in the 
WKB contribution occur when the multicovering index $w$ in (\ref{lim-ns}) takes the value $w=\ell s$, $\ell=1,2,\cdots$. These poles are simple and they appear in the coefficient of 
$\re^{-s \ell {\bf d} \cdot \boldsymbol{T}}$ in the power series. For a given spin content $j_L$, $j_R$ and degree ${\bf d}$, 
the pole in this coefficient has residue
\be
\label{res1}
-{2 d_k \over s^2 \ell^2} (-1)^{(2j_L+2j_R+1) r\ell + s \ell {\bf d} \cdot \boldsymbol{B}} N_{j_L, j_R}^{\bf d} (2j_L+1) (2j_R+1). 
\ee
In the non-perturbative contribution, the poles at $\hbar =2 \pi r/s$ occur when the multicovering takes the values $w=\ell r$, $\ell=1,2,\cdots$, and they appear in the coefficient of the same 
term $\re^{-s \ell {\bf d} \cdot \boldsymbol{T}}$. The corresponding residue is 
\be
\label{res2}
{2 d_k \over s^2 \ell^2} (-1)^{(2j_L+2j_R+1) s \ell + r \ell {\bf d} \cdot \boldsymbol{B}} N_{j_L, j_R}^{\bf d} (2j_L+1) (2j_R+1). 
\ee
Therefore, the condition (\ref{B-prop}) guarantees the cancellation of the poles. In particular, the sum of the two terms in (\ref{combi}) makes sense as a formal 
power series in $Q_i$ and $Q_i^{2\pi/\hbar}$. 

 It turns out that, for the geometry $X_{N-1}$, the B-field 
\be
\boldsymbol{B}=\begin{cases} (0,0, \cdots, 0,0), & N \, \text{even}, \\
(0,0, \cdots, 0,1), & N \, \text{odd},\end{cases}
\ee
satisfies the constraint (\ref{B-prop}) and leads precisely to the insertion of a $(-1)^N$ sign in the last component of ${\bf Q}$, as in (\ref{hatfns}). 
We are now ready to state our conjectural, exact quantization condition for the relativistic Toda lattice. It is given by, 
\be
\label{EQC}
\sum_{j=1}^{N-1} C_{kj} \left\{ {\partial  \over \partial T_j} \widehat F^{\rm NS}\left (\widehat{\boldsymbol{T}}(\hbar), \hbar \right) +
\frac{\hbar}{2\pi}
 {\partial \over \partial T_j} F^{\rm NS, BPS}\left({2 \pi \over \hbar} \widehat{\boldsymbol{T}}(\hbar)+ \pi \ri \boldsymbol{B}, {4 \pi^2  \over \hbar} \right)\right\}= 2 \pi  \left( n_k+{1\over 2} \right),
 \ee
for $ k=1, \cdots, N-1$. The condition (\ref{EQC}) has the following properties:

\vspace{0.3cm}

\begin{enumerate}

\item The quantization condition for given a set of integers ($n_1$, $\dots$, $n_{N-1}$) 
determines the values ($T_1$, $\dots$, $T_{N-1}$) uniquely.
Once they are known, one can easily compute the original eigenvalues ($H_1$, $\dots$, $H_{N-1}$) by the inverse of the quantum mirror map 
(see \eqref{eq:qmirrormap1} in the case of $N=3$, for instance).

\item The second term in the brackets is purely non-perturbative in $\hbar$, so if we expand the l.h.s.  
in a power series around $\hbar=0$, only the first term contributes, and we recover the all-orders WKB quantization condition (\ref{all-WKB}). 

\item Poles cancel, so the l.h.s. is a well-defined power series in $Q_i$ and $Q_i^{2\pi/\hbar}$. The resulting series seems to be convergent in a neighborhood of the large radius limit $T_i \rightarrow +\infty$, 
which corresponds to large Hamiltonians. 

\item In the case $N=2$, (\ref{EQC}) reduces to the exact quantization condition for local $\IP^1 \times \IP^1$, in the form proposed in \cite{wzh}. This quantization 
condition has been successfully tested in \cite{km,hw,ghm,gkmr}. 

\end{enumerate}

In view of these properties, (\ref{EQC}) is a very natural proposal for an exact quantization condition in the relativistic Toda lattice. 
However, the proof of the pudding is in the eating, and we will perform a detailed test of (\ref{EQC}) in the next section. 

There are some further remarks we can do about our exact quantization condition. First of all, as emphasized in \cite{hatsuda} in the case $N=2$, 
the condition (\ref{EQC}) is invariant under the exchange 
\be
\hbar \leftrightarrow {4 \pi^2 \over \hbar}, \qquad \boldsymbol{T} \leftrightarrow {2 \pi \over \hbar} \boldsymbol{T}. 
\ee
This leads to a symmetry in the spectrum under the S-duality transformation of $\hbar$. However, the symmetry is not manifest when the spectrum is parametrized in terms of 
the eigenvalues $H_i$, $i=1, \cdots, N-1$; one has to use the quantum mirror map and relate them to the K\"ahler parameters $\boldsymbol{T}$. This hidden symmetry is probably a manifestation 
of the fact that the relativistic Toda lattice has a ``modular double," which is related to the original one by the S-duality transformation in the Planck constant \cite{kls}.

The self-dual point $\hbar=2 \pi$ is very special. In the quantization of mirror curves, this point has been called the ``maximally supersymmetric case" \cite{ghm}, by analogy with the situation 
in ABJM theory \cite{cgm-abjm}. For this value of $\hbar$, the spectral theory of quantum mirror curves simplifies 
considerably. The same phenomenon occurs here. It is easy to check that, when $\hbar=2 \pi$, the combination (\ref{combi}) becomes
\be
-{\partial F_0^{\rm BPS} \over \partial T_k} + \sum_{l=1}^n T_l {\partial^2 F^{\rm BPS}_0 \over \partial T_k  \partial T_l}, 
\ee
i.e. it involves just the genus zero free energy of the CY manifold. Note that this term is entirely due to the non-perturbative term in (\ref{EQC}). 
It is easy to see that, in the quantum mirror map for $T_1, \cdots, T_{N-1}$, setting $\hbar=2 \pi$ is equivalent to setting $\hbar=0$ and changing $z_N \rightarrow (-1)^N z_N$. 
Therefore, at the self-dual point $\hbar=2 \pi$, $\widehat{\boldsymbol{T}}(\hbar)$ reduces to the {\it conventional} mirror map. In addition, the 
B-field in (\ref{EQC}) is cancelled by the sign coming from the residue (\ref{res2}). At the end of the day, the condition (\ref{EQC}) becomes,
\be
\label{SD}
\sum_{j=1}^{N-1} C_{kj} \left\{ -{\partial F_0 \over \partial T_j} +
 \sum_{l=1}^N T_l {\partial^2 F_0 \over \partial T_j \partial T_l} +8 \pi^2  b_j^{\rm NS} \right\}= 4 \pi^2 \left( n_k +{1\over 2} \right), \qquad 
k=1, \cdots, N-1.
\ee
This simplified quantization condition at the self-dual point involves essentially the 
information appearing in the Bohr--Sommerfeld approximation (the only information coming from the next-to-leading order appears in the coefficients $b_i^{\rm NS}$). In this case, 
our quantization condition is clearly given by a convergent series expansion, since the quantities in the l.h.s. of (\ref{SD}) are governed by a Picard--Fuchs equation 
and they are known to converge in a neighborhood of the large radius point 
$\boldsymbol{T} \rightarrow \infty$. As we will see in an example in the next section, the actual eigenvalues seem to be all in this domain of convergence.

\sectiono{Testing the conjecture}
\label{testing}

\subsection{$N=2$}
Let us now present some detailed tests of the conjecture (\ref{EQC}). We first note that, 
in the case $N=2$, the Hamiltonian of the relativistic Toda lattice is nothing but the operator $\mathsf{O}_{\IF_0}$ considered in \cite{ghm}. To see 
this in some detail, we use canonically conjugate Jacobi coordinates for the relative motion
\be
\mq={1\over {\sqrt{2}}}(\mq_1-\mq_2), \qquad \mm={1\over {\sqrt{2}}}(\mm_1-\mm_2), 
\ee
and we decouple the motion of the center of mass. After doing this, 
the classical Hamiltonian of the $N=2$ relativistic Toda lattice reads
\be
\mH_1= \re^{R \mm/{\sqrt{2}}}+ \re^{-R \mm/{\sqrt{2}}}+ R^2\left( \re^{{\sqrt 2} \mq+ R \mm/{\sqrt{2}}}+  \re^{-{\sqrt 2} \mq- R \mm/{\sqrt{2}}}\right). 
\ee
After the linear canonical transformation 
\be
\mathsf{\xi}= {\sqrt 2} \mq+R \mm/{\sqrt{2}}, \qquad \mathsf{\mu}= -\mm/{\sqrt{2}}, 
\ee
we put the Hamiltonian in the form 
\be
\mH_1= R^2\left(\re^{\mathsf{\xi}} + \re^{-\mathsf{\xi}}\right)+ \re^{R\mathsf{\mu}}+ \re^{-R \mathsf{\mu}}, 
\ee
which is nothing but the quantization of the spectral curve (\ref{rt-sc}) in the case $N=2$ (since we have decoupled the center of mass motion, we have to set $\mu_1+\mu_2=0$.) 
In this case, as expected, the quantum Baxter equation is just the 
original eigenvalue problem for the Hamiltonian, once the center of mass motion has been decoupled.  
Finally, by a further linear transformation,
\be
\mx= \mathsf{\xi}+ 2 \log R, \qquad \my=-R \mathsf{\mu}. 
\ee
we can write the Hamiltonian as 
\be
\label{HO}
\mH_1=\re^{\mx}+ m_{\IF_0} \re^{-\mx} + \re^{\my} + \re^{-\my}, 
\ee
with 
\be
m_{\IF_0}= R^4.
\ee
Note that $\mx, \my$ are Heisenberg operators satisfying 
\be
[\mx, \my] =\ri \hbar, \qquad \hbar=R \hbar_{\rm RT}.
\ee
The Hamiltonian $\mH_1$, written in terms of $\mx$, $\my$, is nothing but the operator $\mathsf{O}_{\IF_0}$. A solution for the spectral problem of this operator was proposed in \cite{ghm}, and later 
reformulated in \cite{wzh} in the form (\ref{EQC}). Therefore, in the case $N=2$, our proposal for the exact quantization condition of the relativistic Toda lattice is backed by the extensive evidence 
for the proposal of \cite{ghm}. 

\subsection{$N=3$}

Let us then consider the case $N=3$. To verify our proposal (\ref{EQC}) in this case, we simply calculate numerically the spectrum of the relativistic Toda lattice, and 
we compare the results with the 
predictions obtained from (\ref{EQC}).  

We first note that the bound states of this system can be labelled by the two quantum numbers $(n_1, n_2)$ appearing in the Bohr--Sommerfeld 
quantization condition (\ref{bs-qc}) and its exact counterpart (\ref{EQC}). There is also a symmetry in the spectrum, 
since the exchange of quantum numbers
\be
(n_1, n_2) \leftrightarrow (n_2, n_1)
\ee
corresponds to the exchange of the two eigenvalues
\be
\label{symH}
(H_1, H_2) \leftrightarrow (H_2, H_1).
\ee

In order to compute the spectrum, the most direct route is to adapt the numerical methods developed in \cite{matsu-plb, matsuyama, isola} for the standard Toda lattice. To do this, we first 
write the Hamiltonians in Jacobi coordinates, so that one can decouple the center of mass motion. These coordinates are given by 
\be
\ba
\zeta_1&={1\over {\sqrt{2}}}(q_1-q_2), \\
\zeta_2&={1\over {\sqrt{2}}}(q_1+q_2-2q_3), \\
\zeta_0&={1\over 3}(q_1+ q_2+q_3). 
\ea
\ee
For the momenta, we have
\be
\ba
p_1&= {1\over {\sqrt{2}}} p_{\zeta_1}+ {1\over {\sqrt{6}}} p_{\zeta_2}+ {1\over 3} p_{\zeta_0},\\
p_2&= -{1\over {\sqrt{2}}} p_{\zeta_1}+ {1\over {\sqrt{6}}} p_{\zeta_2}+ {1\over 3} p_{\zeta_0},\\
p_3&=   -{2\over {\sqrt{6}}} p_{\zeta_2}+ {1\over 3} p_{\zeta_0}. 
\ea
\ee
In order to remove the center of mass movement, we simply set $p_{\zeta_0}=\zeta_0=0$. In this way, we re-express the Hamiltonians $\mH_{1,2}$ in 
terms of two sets of positions and momenta. An appropriate basis for 
the Hilbert space is then given by 
\be
\label{basis}
\langle \zeta_1, \zeta_2 |m_1, m_2 \rangle=\phi_{m_1} (\zeta_1) \phi_{m_2}(\zeta_2), \qquad m_1, m_2=0, 1, 2, \cdots, 
\ee
where $\phi_m(\zeta)$ are the eigenfunctions for a quantum one-dimensional harmonic oscillator. One then 
considers the matrix 
\be
\label{H-matrix}
\langle \ell_1, \ell_2 |\mH_1| m_1, m_2\rangle, 
\ee
which can be diagonalized after truncating the basis to a finite set of $M$ elements. The eigenvalues of this matrix provide an approximation to the eigenvalues of $\mH_1$, which converge to the 
exact values as we increase $M$. The approximate eigenvalues of $\mH_2$ can be obtained by calculating its vacuum expectation values in the approximate eigenfunctions obtained in the diagonalization 
of $\mH_1$. The frequency of the harmonic oscillator providing the basis (\ref{basis}) can be chosen so as to improve the 
convergence (we normalize its mass to be one). A useful formula to compute the matrix elements is the following, 
\be
\langle m | \re^{a x} \re^{b p} |\ell\rangle=2^{m+ \ell \over 2} {\sqrt{m! \ell!}} \, \re^{|z|^2 + \im  \hbar a b /2} z^m \overline z^\ell \sum_{k=0}^{ {\rm min}(m, \ell)} {1\over k! (m-k)! (\ell-k)!} {1\over (2 |z|^2)^{k}}, 
\ee
where $|m\rangle$ is the $m$-th energy eigenstate of the harmonic oscillator, $\omega$ is its frequency, and 
\be
z={1\over 2} \left( {\hbar \over\omega}\right)^{1/2} \left( a+  \im\omega b \right).
\ee

\begin{figure}[tb]
\begin{center}
\begin{tabular}{cc}
%\hspace{-3mm}
\resizebox{70mm}{!}{\includegraphics{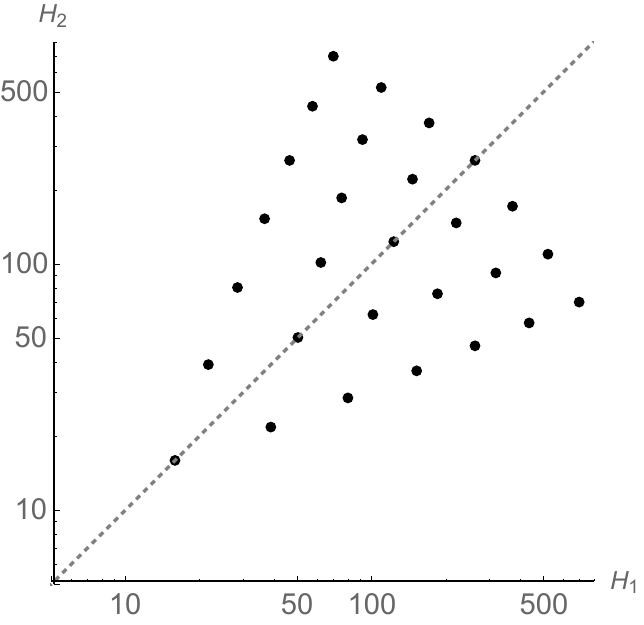}}
\hspace{3mm}
&
\resizebox{65mm}{!}{\includegraphics{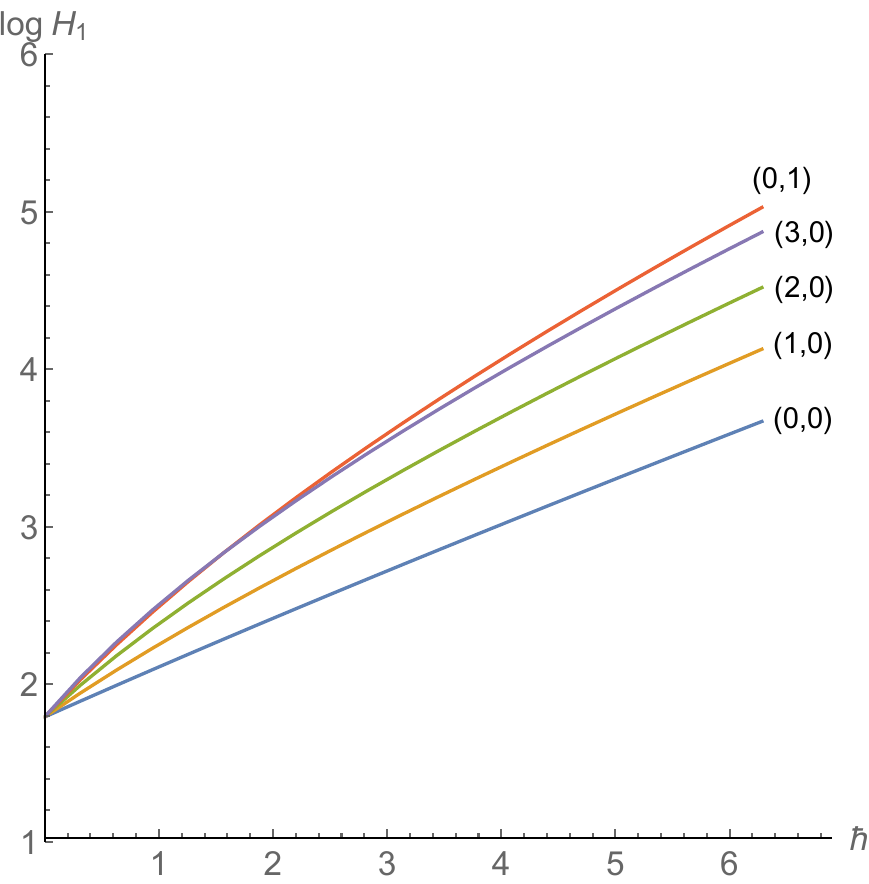}}
%\vspace{5mm}
\end{tabular}
\end{center}
  \caption{(Left) The eigenvalue distribution in the $N=3$ relativistic Toda lattice for $\hbar=\pi$ and $R=1$.
  (Right) The coupling dependence of $\log H_1$ with the quantum numbers $(n_1,n_2)=(0,0), (1,0), (0,1), (2,0), (3,0)$ and $R=1$.
  In the purely classical limit $\hbar \to 0$, all the eigenvalues degenerate to $H_1=H_2=6$.
}
\label{fig:eigenvalues}
\end{figure}

It turns out that this method of computing the eigenvalues is computationally expensive, specially if one wants good numerical precision. One can obtain much more precise 
answers by looking at the quantum Baxter equation, which in the $N=3$ case can be obtained by a direct quantization of the spectral curve
\be
R^3\left( \re^{\xi}+ \re^{-\xi}\right)+\re^{3R \mu/2}- \re^{-3 R \mu/2} -H_1  \re^{R\mu/2}  +H_2  \re^{-R\mu/2}=0.
\ee
After promoting $\xi$ to a differential operator, as in (\ref{xi-dif}), we obtain the operator equation  
\be
\label{baxe}
\left\{ R^3\left(\re^{\xi-R\mu/2}+\re^{-\xi-R\mu/2}\right)+\re^{R\mu}-\re^{-2R\mu}+H_2 \re^{-R\mu} \right\} |\psi\rangle= H_1 |\psi\rangle. 
\ee
It turns out that, if we require $\psi(\mu)$ to be in $L^2(\IR)$, this equation only has a solution if the values of $H_1$, $H_2$ are simultaneously quantized. In practice, this means that the numerical 
diagonalization process leading to an eigenvalue $H_2$ will only converge if $H_1$ takes special values. 
We can also exploit the symmetry (\ref{symH}) to search for pairs of eigenvalues which lead to an admissible solution to (\ref{baxe}). 
For example, the ground state, 
with quantum numbers $n_1=n_2=0$, has $H_1=H_2$, so one looks for admissible solutions of (\ref{baxe}) 
in which $H_1$ is equal to $H_2$. 
Since this is a one-dimensional problem, we can obtain much more precision for the spectrum than with the direct diagonalization 
of the two-dimensional problem. 
We have verified that both methods lead to the same results, taking into account numerical precision. For example, 
for $\hbar =\pi$ and $R=1$, a time-consuming calculation of the eigenvalues of 
(\ref{H-matrix}) gives, for the ground state, 
\be
H_1=H_2=15.8137841...
\ee
while the quantum Baxter equation gives an improved precision without much time
\be
H_1=H_2=15.8137841054...
\ee
In this way, we can calculate the eigenvalues numerically.
In the left of figure~\ref{fig:eigenvalues}, we show some of the allowed eigenvalues of $(H_1,H_2)$ for $\hbar=\pi$ and $R=1$.
The distribution is symmetric with respect to the line $H_1=H_2$, as expected from (\ref{symH}).
In the right of figure~\ref{fig:eigenvalues}, the dependence of $\log H_1$ on $\hbar$ is shown for several low quantum numbers.

Let us now address the conjectural, exact quantization condition (\ref{EQC}). In order to implement it, we have to calculate the NS free energy for the CY geometry $X_2$. First of all, we note that the 
prepotential, which is the leading term in the NS free energy, can be computed with local mirror symmetry. In this case, the charge vectors are given by 
\be
\ba
e_1&=(0,0,1,-2,1,0),\\
e_2&=(0,0,0,1,-2,1),\\
e_3&= ( 1,1,-1,0,0,-1),\\
\ea
\ee
and the Batyrev coordinates read, 
\be
 z_1= {H_2 \over H_1^2}, \qquad  z_2= {H_1 \over H_2^2}, \qquad z_3=R^6.
\ee
The periods of the above CY can be calculated by using standard technology (see for example \cite{hosono}). The fundamental period is given by 
\be
\omega_0( \boldsymbol{\rho})=\sum_{l,m,n\ge 0} c(l,m,n; \boldsymbol{\rho})z_1^{l+ \rho_1} z_2^{m+ \rho_2} z_3^{n + \rho_3},
\ee
where
\be
\ba
c(l,m,n; \boldsymbol{\rho})&={1\over \Gamma (n+\rho_3+1)^2 \Gamma
   (-n+l+\rho_1-\rho_3+1) \Gamma (-n+m+\rho_2-\rho_3+1) }\\
    & \times{1\over  \Gamma (l-2 m+\rho_1-2 \rho_2+1) \Gamma (-2 l+m-2 \rho_1+\rho_2+1)}.
   \ea
 \ee
From here we can form the building blocks for the periods. We define, as usual, 
\be
\omega_i={\partial \omega_0 \over \partial \rho_i}\biggl|_{{\boldsymbol \rho}=0}, \qquad \omega_{ij}={\partial ^2\omega_0 \over \partial \rho_i \partial \rho_j}\biggl|_{{\boldsymbol \rho}=0}.
\ee
The first derivative defines the mirror map, 
\be
-T_i = \omega_i, \qquad i=1,2,3. 
\ee
We have, explicitly,
\be
\ba
-T_1(z_1, z_2, z_3)&= \log(z_1)+2z_1 -z_2 + 3z_1^2 -{3 z_2^2 \over 2}+{20 z_1^3 \over 3} - 2 z_1^2 z_2+ z_1 z_2^2 - {10 z_2^3 \over 3}+ \cdots \\
T_2(z_1, z_2,z_3)&= T_1(z_2, z_1, z_3)  \\
-T_3(z_3)&=\log(z_3). 
\ea
\ee
As expected, $z_3$ is a parameter and its mirror map is algebraic. The genus zero free energy is defined by the equations, 
\be
\ba
{\partial F_0 \over \partial T_1}&=  \omega_{11} + \omega_{12} + {1\over 2} \omega_{22} + {2 \over 3} \omega_{13}+ {1\over 3} \omega_{23}+ {2 \pi^2 \over 3}, \\
{\partial F_0 \over \partial T_2}&={1\over 2}  \omega_{11} + \omega_{12} +  \omega_{22} + {1\over 3} \omega_{13}+ {2\over 3} \omega_{23}+ {2 \pi^2 \over 3}, 
\ea
\ee
and one finds
\be
\ba
F_0( {\bf Q})&=F_0^{\rm pert}({\bf Q}) -2 \left( {\rm Li}_3 (Q_1) + {\rm Li}_3 (Q_2) + {\rm Li}_3(Q_1 Q_2) \right)\\
&+ Q_B+ 3 (Q_1 + Q_2) Q_B+  (5Q_1^2 + 4 Q_1 Q_2 +5 Q_2^2) Q_B+ \cdots, 
\ea
\ee
where
\be
\label{qb-def}
Q_i= \re^{-T_i}, \quad i=1,2,3, \qquad Q_B= Q_1 Q_2 Q_3, 
\ee
and
\be
F_0^{\rm pert}({\bf Q})={T_1^3\over 3} +{T_2^3 \over 3} +{1\over 2} \left( T_1^2 T_2 + T_1 T_2^2 \right) + {1\over 3} (T_1^2  + T_2^2)T_3+ {1\over 3} T_1 T_2 T_3.
\ee
With this information, we can already test our exact quantization condition in the self-dual case $\hbar=2 \pi$, given in (\ref{SD}). To do this, we compute successive 
approximations to the l.h.s. of (\ref{SD}): we consider the large radius expansion of the prepotential and we truncate it to higher and higher total degrees. We then solve numerically the 
quantization condition by using the truncated prepotential. We present some of the results in tables \ref{sd-one} and 
\ref{sd-two}. The agreement is excellent. We have tested the agreement for other values of $R$ and/or other values of the integers, and we found again full agreement, at the level of numerical 
precision that we achieved.

%%%%%
 \begin{table}[t] 
\centering
   \begin{tabular}{l l }
  \\
Order& $H_1$ \\
\hline
 1&  \underline{39.1}104429532554969\\ 
 5&   \underline{39.1678}325080157194\\
 15&    \underline{39.16781907627}95935\\ 
 23 & \underline{39.1678190762768699}\\
 \hline
Numerical value &
                         $39.1678190762768699$   \\
\end{tabular}
\\
\caption{ The eigenvalue of the Hamiltonians $H_1=H_2$ for the quantum numbers $(n_1, n_2)=(0,0)$ in the $N=3$ relativistic Toda lattice with $R=1$ and $\hbar=2 \pi$, 
as obtained from the quantization 
condition (\ref{SD}). The order denotes the total degree in the moduli $z_i$. As we keep more and more terms in the series in the l.h.s. of (\ref{SD}), 
we quickly approach the eigenvalue obtained by numerical methods.}
 \label{sd-one}
\end{table}

%%%%%
 \begin{table}[t] 
\centering
   \begin{tabular}{l l l}
  \\
Order& $H_1 $  & $H_2$\\
\hline
 1 &   \underline{61}.7259698869968 & \underline{152}.405034932001\\ 
 6 &   \underline{61.9664}326975787& \underline{152.35967}6263718\\ 
 12&    \underline{61.966419006}6106& \underline{152.35967200}0995\\ 
 18 & \underline{61.9664190064911} & \underline{152.359672001068}\\
 \hline
Numerical value &
                         $61.9664190064911$  & $152.359672001068 $ \\
\end{tabular}
\\
\caption{ The eigenvalues of $H_1$ and $H_2$ for the quantum numbers $(n_1,n_2)=(1,0)$ in the $N=3$ relativistic Toda lattice with $R=1$ and $\hbar=2 \pi$, 
as obtained from the quantization 
condition (\ref{SD}). The last line displays the eigenvalue obtained by numerical methods.}
 \label{sd-two}
\end{table}

In order to test the full conjecture (\ref{EQC}), for arbitrary values of $\hbar$, we need the quantum mirror map and the NS free energy. 
As we mentioned in the previous section, the quantum mirror map can be computed by quantization of the mirror curve, by using 
(\ref{qc-gen}). Different representations of the curve, related by canonical transformations, lead to the different, independent mirror maps. One finds, by quantizing 
(\ref{genmc}), 
\be
V(X)+{ z_1 z^2_2 z_3 q^{3/2} X^3 \over V(q X)} + z_1 z_2^2 X^3+z_2 X^2 + X +1=0, 
\ee
where
\be
V(x)={\psi(x-\ri \hbar) \over \psi(x)}, \qquad X=\re^x. 
\ee
In this way we obtain 
\be
\ba
-\Pi(z_1, z_2, z_3;\hbar)&= {1\over 3} \log(z_1) +{2\over 3} \log(z_2) + {\rm Res}_{X=\infty}\left( {\log (V(X)) \over X }\right)\\
&= {1\over 3} \log(z_1) +{2\over 3} \log(z_2) + z_2+ {3 z_2^2 \over 2} - z_1 z_2^2 + {10 z_2^3 \over 3}-4 z_1 z_2^3  \\
&-\left( q^{1/2}+{1\over q^{1/2}} \right) z_1 z_2^2 z_3+\frac{35 z_2^4}{4}+\cdots
\ea
\ee
The quantum period $T_1(\hbar)$ is then given by 
\be
T_1(\hbar)=- \Pi(z_1, z_2, z_3;\hbar)+2 \Pi(z_2, z_1, z_3;\hbar), 
\label{eq:qmirrormap1}
\ee
and $T_2(\hbar)$ is obtained by exchanging $z_1 \leftrightarrow z_2$. In addition, $T_3(\hbar)=T_3$. As noted in the last section, setting $\hbar=2 \pi$ is equivalent to setting $\hbar=0$ and changing 
the sign $z_3 \rightarrow -z_3$. This implies that, in the self-dual case, the quantum mirror map ${\widehat {\boldsymbol{T}}}(\hbar)$ becomes the classical mirror map. 

In order to compute the NS free energy, we compute the full, refined free energy of the geometry $X_2$ and then take the NS limit. 
Expressions for the refined free energy can be found in \cite{ikv,taki}. In order to 
write it explicitly, we need some ingredients. The first one is given by
\be
Z_\mu(t, q)= \prod_{(i,j) \in \mu} \left(1- t^{\mu_j^t-i+1} q^{\mu_i-j}\right)^{-1}, 
\ee
where $\mu$ is a partition or Young tableau, and the parameters $q$, $t$ were introduced in (\ref{qt}). 
The second one depends on two partitions $\mu$, $\nu$, and an extra parameter $Q$. It is given by 
\be
R_{\mu \nu}(Q)= \prod_{i,j=1}^\infty {1- Q t^{j-1/2} q^{i-1/2} \over 1-  Q t^{j-1/2- \mu_i} q^{i-1/2-\nu_j}}. 
\ee
It is easy to see that the product gets truncated and only a finite number of factors get involved. We also introduce, for a given partition $\mu$, 
the quantities
\be
\ba
|\mu|&=\sum_i \mu_i, \\
\Vert \mu \Vert^2&= \sum_i \mu_i^2,\\
\kappa_\mu&= \sum_i \mu_i(\mu_i-2i+1),
\ea
\ee
and the refined framing factor
\be
f_\mu= (-1)^{|\mu|}\left( {t \over q} \right)^{\Vert \mu^t \Vert^2/2} q^{-\kappa_\mu/2}, 
\ee
where $\mu^t$ denotes the transposed partition in which one exchange rows and columns of the corresponding Young diagram. 
The building block of the partition function is 
\be
\ba
Z_{\mu^1, \mu^2, \mu^3}&=q^{\sum_{i=1}^3 \Vert \mu^i\Vert^2/2} t^{\sum_{i=1}^3 \Vert \mu^{i,t}\Vert^2/2} \prod_{i=1}^3Z_{\mu^i} (t,q) Z_{\mu^{i,t}}(q,t)
R_{\mu^{1,t},\mu^2}\left( {\sqrt{t\over q}} Q_1\right)R_{\mu^{1,t},\mu^2}\left( {\sqrt{q\over t}} Q_1\right) \\
& \times R_{\mu^{1,t},\mu^3}\left( {\sqrt{t\over q}} Q_1 Q_2 \right) 
R_{\mu^{1,t},\mu^3}\left( {\sqrt{q\over t}} Q_1 Q_2\right) R_{\mu^{2,t},\mu^3}\left( {\sqrt{t\over q}} Q_2 \right)R_{\mu^{2,t},\mu^3}\left( {\sqrt{q\over t}} Q_2\right).
\ea
\ee
Then, the total partition function of an $A_2$ fibration over $\IP^1$ with Chern--Simons invariant $m$ (see \cite{ikp2}) is given by 
\be
\ba
Z_m \left( {\bf Q};q,t \right)&= Z_{\rm p}(Q_1, Q_2; q,t) \sum_{\mu^i}  f_{\mu^1}^{-m-2} f_{\mu^2}^{-m} f_{\mu^3}^{-m+2}  Z_{\mu^1, \mu^2, \mu^3}\\
& \cdot (-Q_B)^{\sum_{i=1}^3 |\mu^i|}  Q_1^{(m+1)|\mu^1|} 
Q_2^{(m-1)(1-\delta_{m,0})(|\mu^1|+|\mu^2|) +\delta_{m,0} |\mu^3|}, 
\ea
\ee
where
\be
Z_{\rm p}(Q_1, Q_2; q,t)=\exp\left[ \sum_{w\ge 1} {1\over w} { (q/t)^{w/2}+ (t/q)^{w/2} \over \left(q^{w/2}- q^{-w/2}\right) \left( t^{w/2}- t^{-w/2}\right)} \left(Q_1^w+ Q_2^w + Q_1^w Q_2^w\right) \right].
\ee
In the case $q=t$, which corresponds to the standard topological string, this expression agrees with the result obtained in \cite{ikp2}. $Q_B$ 
is defined in (\ref{qb-def}). From now on we will focus on the 
geometry with $m=0$, which is the relevant one for our purposes, and we will denote its partition function simply by $Z({\bf Q}; q,t)$. One finds, explicitly, 
\be
\ba
\ri F^{\rm NS, \, BPS }({\bf Q}, \hbar)&=\sum_{w\ge 1}  {1\over w^2} {q^w+1 \over q^w-1} \left( Q^w_1+ Q^w_2 + Q_1^w Q_2^w \right) -\sum_{w\ge 1} {1\over w^2} {Q_B^w \over q^{w/2}-q^{-w/2}}\\
& -{q+ q^{-1}+1 \over q^{1/2}-q^{-1/2}}(Q_1 + Q_2) Q_B- {q+ q^{-1}+2 \over q^{1/2}-q^{-1/2}} Q_1 Q_2 Q_B + \cdots, 
\ea
\ee 
where some of the poles at (\ref{rat-hbar}) are manifest. 

%%%%%
 \begin{table}[t] 
\centering
   \begin{tabular}{l l l}
  \\
Order& $\hbar=\pi $  & $\hbar=3 \pi$\\
\hline
 1 &     \underline{15}.7049125387& \underline{93.57}56026639\\ 
 5 &     \underline{15.813}6736201& \underline{93.57006}57722\\ 
 10&    \underline{15.813784}0616& \underline{93.5700660274}\\ 
 15 &   \underline{15.8137841054} & \underline{93.5700660274}\\
 \hline
Numerical value &
                         $15.8137841054$  & $93.5700660274$ \\
\end{tabular}
\\
\caption{ The eigenvalue of $H_1=H_2$ for the quantum numbers $(n_1, n_2)=(0,0)$ in the $N=3$ relativistic Toda lattice with $R=1$ and $\hbar=\pi, 3 \pi$, 
as obtained from the quantization 
condition (\ref{SD}). The last line displays the eigenvalue obtained by numerical methods.}
 \label{table-hen}
\end{table}

With these explicit results, we can already test the conjecture for some other values of $\hbar$. In Table \ref{table-hen} we show some results for the ground state energy eigenvalues. 
Precise computations for general $\hbar$ are more demanding, but we find again a remarkable agreement between the predictions of (\ref{EQC}) and the numerical diagonalization. 

One could think that the non-perturbative correction in (\ref{EQC}) is only required when $\hbar$ is of the form (\ref{rat-hbar}). However, 
this set of poles is dense in the positive real axis. Therefore, it is difficult to make sense of (\ref{lim-ns}) for any positive, real $\hbar$, even as a formal power series, 
since there will be infinitely many coefficients in the 
series where $\hbar$ will be arbitrarily close to a pole. If we ignore this issue, and insist on using the quantization condition (\ref{all-WKB}) and 
the expression (\ref{lim-ns}) for values of $\hbar$ which are not of the form (\ref{rat-hbar}), we simply get incorrect results for the spectrum. 
For example, when $R=1$ and $\hbar=3, 10$, (\ref{all-WKB}) gives
$H_1=15.203...$, and $H_1=108.475...$, respectively. These values seem to be stable, at least when working up to degree $12$ in the expansion of (\ref{lim-ns}). 
However, they do not agree with a numerical calculation of the ground state energy. The values obtained with the corrected quantization condition (\ref{EQC}) 
are shown in table \ref{table-irrat}. They are in agreement with the numerical result. 

%%%%%
 \begin{table}[t] 
\centering
   \begin{tabular}{l l l}
  \\
Order& $\hbar=3 $  & $\hbar=10$\\
\hline
 1 &     \underline{15}.0088290209& \underline{109.4}95054032\\ 
 5 &     \underline{15.162}0665172& \underline{109.44}3122441\\ 
 10&    \underline{15.1622789}237& \underline{109.44299472}6\\ 
 15 &   \underline{15.1622789846}& \underline{109.442994727}\\
 \hline
Numerical value &
                          $15.1622789846$  & $109.442994727$ \\
\end{tabular}
\\
\caption{ The eigenvalue of $H_1=H_2$ for the quantum numbers $(n_1, n_2)=(0,0)$ in the $N=3$ relativistic Toda lattice with $R=1$ and $\hbar=3,10$, 
as obtained from the quantization 
condition (\ref{SD}). The last line displays the eigenvalue obtained by numerical methods.}
 \label{table-irrat}
\end{table}

\sectiono{The four-dimensional limit}
In this section we will consider the four-dimensional limit $R \to 0$.
As mentioned before, in this limit the relativistic Toda lattice reduces to the standard Toda lattice.
What happens if one takes the limit $R \to 0$ in our exact quantization conditions \eqref{EQC}? First of all, one has to specify a scaling regime for the 
K\"ahler parameters. From previous studies of this limit \cite{ikp,ikp2,ek}, it is known that 
\be
T_i = R a_{i,i+1}, \qquad i=1, \cdots, N-1,  
\ee
where 
\be
a_{i,i+1}= a_i-a_{i+1}
\ee
and the $a_i$ are parameters for the Coulomb branch of the 4d supersymmetric gauge theory. In particular, we have that
\be
T_l+T_{l+1} +\cdots + T_n = R \left(a_l - a_n \right) = Ra_{l,n}. 
\ee

It turns out that there are four different terms to consider on the l.h.s. of the exact quantization condition (\ref{EQC}), with different limits as $R\rightarrow 0$. 
The first piece comes from the perturbative part of the NS free energy, (\ref{pertNS}). Each pair $1\le l< n\le N-1$ leads to a divergent term
\be
\label{div-pert}
-{\pi^2 \over 3 \hbar_{\rm RT} R} - {2 \over \hbar_{\rm RT}} \log(R) a_{l,n}. 
\ee
The second piece comes from the part of the BPS free energy (\ref{lim-ns}) which depends on $Q_N$. This piece goes straightforwardly into 
the instanton part of the 4d quantization condition, as it can be checked with the techniques of \cite{ikp,ikp2}. 
The third piece comes from the part of the S-transformed BPS free energy which depends on $Q_N$. This piece vanishes as $R\rightarrow 0$, since 
\be
Q_N^{2 \pi/\hbar} = R^{4 \pi N /\hbar_{\rm RT} R}. 
\ee
Finally, the fourth piece involves the part of the BPS free energy which does not depend on $Q_N$, plus its S-transform. It is a sum of terms of the form  $f(Q_{l,n}, \hbar)$,  
 where
\be
Q_{l,n}= \re^{-T_l -\cdots -T_n}= \re^{-R a_{l,n}}, 
\ee
and the function $f$ is given by
\be
f(Q, \hbar)= \sum_{w=1}^\infty {1\over w} \cot\left( {\hbar w \over 2} \right) Q^w+\sum_{w=1}^\infty \frac{1}{w} \cot \( \frac{\hbar_{\rm D} w}{2} \) Q_{\rm D}^w. 
\label{eq:f-1}
\ee
In this equation, we have denoted
\be
Q_{\rm D}=Q^{2 \pi/\hbar},\qquad
\hbar_{\rm D}=\frac{4\pi^2}{\hbar}.
\ee
This function can be also expressed in terms of Faddeev's quantum dilogarithm $\fad(x)$, as follows
\be
\im f(Q, \hbar) = \log \fad \left( -{T \over 2 \pi \mb} - c_{\mathsf{b}}\right)+ \log \fad \left( -{T \over 2 \pi \mb} +c_{\mathsf{b}}\right), 
\label{eq:f-2}
\ee
where
\be
Q=\re^{-T}, \qquad c_{\mathsf{b}}= {\im \over 2} \left( \mb +\mb^{-1} \right), 
\ee
and $\mb$ is related to $\hbar $ by 
\be
\mb^2={\hbar \over 2 \pi}. 
\ee
Our conventions for Faddeev's quantum dilogarithm are e.g. as in \cite{ak}. 
By manipulations similar to those used in \cite{ho}, one can find a useful integral expression for $f(Q, \hbar)$:
\be
\label{fQex}
f(Q, \hbar)= \frac{2}{\hbar} \Li_2(Q)+\frac{2}{\pi} {\rm Re} \biggl[
\int_0^{\infty \re^{\pm \ri 0}} \rd x \frac{\hbar Q(\cosh \hbar x-Q)}{(1-Q \re^{\hbar x})(1-Q \re^{-\hbar x})}
\log(1-\re^{-2\pi x}) \biggr] . 
\ee
In the integral on the r.h.s., the integrand has poles at $x=\pm T/\hbar$.
If both $T$ and $\hbar$ are real, one of these poles is located on the positive real axis. 
We can avoid this pole by deforming the contour above or below the real axis, and the two choices correspond to the 
integration limits $\infty \re^{\pm \ri 0}$ in (\ref{fQex}). We keep just the real part of the resulting integral  
(equivalently, we use a principal part prescription). The value of the real part does not depend on the choice of the deformation.

We can now study the limit $R \rightarrow 0$ of the function, after setting $T= R a$. The integrand in (\ref{fQex}) has the limit
\be
\frac{\hbar Q(\cosh \hbar x-Q)}{(1-Q \re^{\hbar x})(1-Q \re^{-\hbar x})}
\log(1-\re^{-2\pi x})
=-\frac{z}{x^2-z^2}\log(1-\re^{-2\pi x})+{\CO}(R),
\ee
where
\be
z=\frac{a}{\hbar_\text{RT}}.
\ee
Let us define the integral 
\be
\CI (z)= -\frac{1}{\pi}  {\rm Re} \left[\int_0^{\infty \re^{\pm \ri 0}} \rd x \, \frac{1}{x^2-z^2} \log(1-\re^{-2\pi x}) \right], 
\label{eq:S-pm-g}
\ee
where we use again a deformation of the contour around the pole. 
Using an integral representation of the logarithm of the gamma function:
\be
\log \Gamma(z)=2\int_0^\infty \rd x \frac{\tan^{-1} (x/z)}{\re^{2\pi x}-1}+\frac{\log 2\pi}{2}
+\( z-\frac{1}{2} \) \log z -z, \quad \real z>0,
\label{eq:logGamma}
\ee
it turns out that the integral \eqref{eq:S-pm-g} can be evaluated as%
\footnote{
In order to go from \eqref{eq:I-Gamma} to \eqref{eq:S-pm-g}, we need to replace $z \to \ri z$ in \eqref{eq:logGamma}.
Strictly, this does not satisfy the condition in \eqref{eq:logGamma} for $z \in \mathbb{R}$.
This corresponds to the fact that the integrand of \eqref{eq:S-pm-g} has poles at $x=\pm z$,
and one has to deform the contour.
However, one can check numerically that the equality of \eqref{eq:S-pm-g} and \eqref{eq:I-Gamma}
indeed holds in the end.
} 
\be
\CI(z)=1-\log z-\frac{\pi}{4z}-\frac{\ri}{2z} \log \frac{\Gamma(1+\ri z)}{\Gamma(1-\ri z)}. 
\label{eq:I-Gamma}
\ee
This result can be understood as a Borel resummation of the following asymptotic expansion around $\hbar_\text{RT}=0$
(recall that $z=a/\hbar_\text{RT}$):
\be
\CI(z) \sim \sum_{n=1}^\infty (-1)^n \frac{B_{2n}}{2n(2n-1)} \frac{1}{z^{2n}} ,\qquad z \to \infty.
\ee
This is closely related to the asymptotic expansion of the function $\gamma_{\im \hbar}(x)$ which is used in \cite{no} to define the perturbative 
part of the instanton partition function. When $\hbar$ is real, as we are assuming here, the asymptotic series above is not Borel summable and there is a non-perturbative ambiguity. Therefore, 
one needs to fix it by using additional information. In our case, this information comes from our starting point, \eqref{eq:f-1} or \eqref{eq:f-2}, and leads 
to the integral representation \eqref{eq:S-pm-g}, which can be regarded as a particular choice of Borel resummation. 

Now, it is easy to see that the limit as $R \rightarrow 0$ of $f(Q, \hbar)$ is given by 
\be
f(Q, \hbar)= {\pi^2 \over 3 \hbar_{\rm RT} R} + {2\over \hbar_{\rm RT}} a \log R + {2\over \hbar_{\rm RT}} a \left( \log a -1\right)
 + {2  a \over \hbar_{\rm RT}} \CI\left( \frac{a}{\hbar_\text{RT}}\right)+\CO(R). 
\ee
The divergent terms appearing in this expression as $R\rightarrow 0$ cancel exactly against the divergent terms (\ref{div-pert}) coming from the perturbative part. We then find that 
the 4d quantization condition has an instanton piece, and a ``perturbative" piece involving an appropriate sum of terms of the form 
\be
{2 \over \hbar_{\rm RT}} \gamma_{4d}\left(a_{l,n}, \hbar_{\rm RT}\right), 
\ee
where
\be
\label{gf4d}
\gamma_{4d}\left(a, \hbar_{\rm RT}\right)=a \left( \log a -1\right) + a  \CI\left( \frac{a}{\hbar_\text{RT}}\right). 
\ee
As a consequence, we conclude that our exact quantization conditions in the limit $R \to 0$ are 
exactly equivalent to the ones proposed in \cite{ns}, and the four-dimensional Planck constant is $\hbar_{\rm RT}$. 
The fact that the quantization conditions in \cite{ns} give the exact spectra in the non-relativistic Toda chain
was also confirmed in \cite{kt} from the perspective of integrable systems.
Note, however, that in the four-dimensional calculation we have to use the function (\ref{gf4d}) as the building block for the ``perturbative" piece, and the prescription to 
deal with this function is inherited from the 5d function (\ref{eq:f-2}), which does include non-perturbative corrections in $\hbar$.

We can check this conclusion quantitatively.
Let us start with the non-relativistic Toda Hamiltonian
\be
\mH_\text{Toda}(\mq_1,\mm_1;\ldots;\mq_{N},\mm_{N}) = \sum_{n=1}^N \left( {\mm_n^2\over 2} + \re^{\mq_n-\mq_{n+1}} \right).
\ee
For simplicity, we here focus on the case of $N=2$.
As in the relativistic case, after modding out the motion of the center of mass, we obtain the Hamiltonian,
\be
\label{mmathieu}
\mH_\text{Toda}^{N=2}=\mm^2+2\cosh \mq,
\qquad [\mq, \mm]=\ri \hbar, 
\ee
where
\be
\mq=\mq_1-\mq_2,\qquad \mm=\frac{\mm_1-\mm_2}{2}.
\ee
(\ref{mmathieu}) is well-known as the Schr\"odinger operator of the modified Mathieu potential.
We want to solve the quantum eigenvalue problem of the modified Mathieu equation for \textit{real} $\hbar$,
\be
%\mH_\text{Toda}^{N=2} \ket{\psi}=E \ket{\psi}.
\ba
\left( -\hbar^2 \frac{\rd^2}{\rd q^2} +2\cosh q\right) \psi(q)=E \psi (q).
\ea
\ee
We here confirm that this problem is indeed solved by the quantization condition in \cite{ns}.
The quantization condition proposed in \cite{ns} takes the form
\be
\frac{\pd {\cal F}_\text{NS}}{\pd a}=2\pi \hbar \left( n+\frac{1}{2} \right),\qquad
n=0,1,2,\dots.
\label{eq:NS-QC}
\ee
The NS free energy ${\cal F}_\text{NS}$ consists of two pieces
\be
\ba
{\cal F}_\text{NS}(a;\hbar)={\cal F}_\text{NS}^\text{pert}(a;\hbar)+{\cal F}_\text{NS}^\text{inst}(a;\hbar).
\ea
\ee
In the current case, this free energy is computed by the Nekrasov partition function in the 4d pure SU(2)
super Yang-Mills theory. 
The perturbative part can be expressed in terms of the function (\ref{gf4d}), as discussed above, 
\be
\ba
\frac{\pd {\cal F}_\text{NS}^\text{pert}}{\pd a}& =-4a \log \left(\Lambda^2 \right)+ 4 \gamma_{\rm 4d}(2a, \hbar)\\
&=
-4a \log \left( \frac{\Lambda^2}{\hbar^2} \right)
-2\ri \hbar \log \frac{\Gamma(1+\frac{2\ri a}{\hbar} )}{\Gamma(1-\frac{2\ri a}{\hbar} )}
-\pi \hbar,
\ea
\ee
and we have simply reinstated the dependence on the scale $\Lambda$. The instanton expansion is given by, 
\be
{\cal F}_\text{NS}^\text{inst}(a;\hbar)=\sum_{n=1}^\infty \Lambda^{4n} {\cal F}_\text{NS}^{(n)}(a;\hbar), 
\label{eq:F-NS}
\ee
where the explicit forms up to $n=3$ are given by
\be
\ba
{\cal F}_\text{NS}^{(1)}(a;\hbar)&=-\frac{2}{4 a^2+\hbar ^2},\\
{\cal F}_\text{NS}^{(2)}(a;\hbar)&=-\frac{20 a^2-7 \hbar ^2}{4 \left(a^2+\hbar ^2\right) \left(4 a^2+\hbar ^2\right)^3} ,\\
{\cal F}_\text{NS}^{(3)}(a;\hbar)&=-\frac{4 \left(144 a^4-232 a^2 \hbar ^2+29 \hbar ^4\right)}{3 \left(a^2+\hbar ^2\right) \left(4 a^2+\hbar ^2\right)^5 \left(4 a^2+9 \hbar ^2\right)} .
\ea
\label{eq:su2-inst}
\ee
These are understood as the all-order $\hbar$-resummation in the instanton expansion.
An important observation is that the series \eqref{eq:F-NS} is \textit{convergent}, as pointed out in \cite{ok}.
It is also important to notice that the energy $E$ and the modulus $a$ are related by the quantum mirror map.
In particular, its inverse relation is known as the Matone relation \cite{matone}:
\be
E=-\frac{\Lambda}{4} \frac{\pd {\cal F}_\text{NS}}{\pd \Lambda}=a^2-\frac{\Lambda}{4} \frac{\pd {\cal F}_\text{NS}^\text{inst}}{\pd \Lambda}.
\ee
Therefore, if the discrete spectrum for $a$ is known from \eqref{eq:NS-QC}, then one can recover the energy
eigenvalues $E$.

Another interesting aspect is the relation to a TBA-like system.
As proposed in \cite{ns}, the instanton part of the NS free energy is completely determined by TBA-type integral equations.
In the non-relativistic Toda lattice, the TBA equation is 
\be
\varphi(x)=-\int_{-\infty}^\infty \frac{\rd y}{2\pi} K(x-y) \log (1+\Lambda^4 Q(y) \re^{-\varphi(y)}),
\label{eq:TBA}
\ee
where
\be
\ba
K(x)=\frac{2\hbar}{x^2+\hbar^2}, \qquad
Q(x)=\prod_{\ell=1}^N \frac{1}{(x-a_\ell-\ri \hbar/2)(x-a_\ell+\ri \hbar/2)}.
\ea
\ee
The TBA equation uniquely fixes the unknown function $\varphi(x)$.
For the solution, the NS free energy is given as the so-called Yang-Yang potential
\be
{\cal F}_\text{NS}^\text{inst}=-\frac{\hbar}{2\pi}
\int_{-\infty}^\infty \rd x \left[ -\frac{1}{2} \varphi(x) \log(1+\Lambda^4 Q(x) \re^{-\varphi(x)})
+\Li_2(-\Lambda^4 Q(x) \re^{-\varphi(x)} ) \right].
\ee
One can check that the small $\Lambda$ expansion for $N=2$ reproduces the instanton expansion \eqref{eq:F-NS} and \eqref{eq:su2-inst}.
The same integral equation was derived from the Baxter equation in the Toda lattice \cite{kt}.

Now, we compare the spectrum obtained by the NS quantization condition \eqref{eq:NS-QC}
with the true one when $N=2$.
We do so in two ways.
One is to use the instanton expansion \eqref{eq:F-NS}.
The other is to directly solve the TBA equation \eqref{eq:TBA} numerically.
This is done by the standard iterative method. To solve the quantization condition by iteration, we need $\pd {\cal F_\text{NS}}/\pd a$ rather than the
free energy itself. There is an integral expression of the derivative
\be
\frac{\pd {\cal F}_\text{NS}^\text{inst}}{\pd a}
=\frac{\hbar}{2\pi} \int_{-\infty}^\infty \rd x \, \widetilde{K}(x) \log(1+\Lambda^4 Q(x) \re^{-\varphi(x)} ),\quad
\ee
where
\be
\widetilde{K}(x)=\frac{\rd}{\rd x} \log \frac{(x+a+\ri \hbar/2)(x+a-\ri \hbar/2)}{(x-a+\ri \hbar/2)(x-a-\ri \hbar/2)}.
\ee
Similarly, in the Matone relation, we can use an expression
\be
\frac{\Lambda}{4} \frac{\pd {\cal F}_\text{NS}^\text{inst}}{\pd \Lambda}
=-\frac{\hbar}{2\pi} \int_{-\infty}^\infty \rd x \, \log(1+\Lambda^4 Q(x) \re^{-\varphi(x)} ).
\ee
We have checked that all these expressions correctly reproduce the instanton expansion for $N=2$. 

We can now compare the eigenvalues obtained by these two methods\footnote{For the comparison, we have to set $\Lambda=1$.} 
with the ones obtained from the numerical diagonalization of the matrix elements
of the Hamiltonian in a harmonic oscillator basis. This is completely in parallel with the computation that was done in the relativistic case.
We show the results for $\hbar=1/2$ and $\hbar=1$ in Tables~\ref{tab:spectrum1} and \ref{tab:spectrum2}, respectively.
It is clear that the spectra obtained from the condition \eqref{eq:NS-QC}, where ${\cal F}_\text{NS}$ is given by the 
instanton partition function, go to the true eigenvalues when more and more instanton corrections are included.
The numerical evaluation of the TBA equation, which includes all the instanton corrections, shows
a remarkable agreement, as expected from the argument in \cite{kt}. We have also tested in detail both methods in the case of 
the standard Toda lattice with $N=3$, with similar agreement.  

\begin{table}[tb]
\begin{center}
  \begin{tabular}{cccc}\hline
	Instanton number & $E_0$ & $E_1$ & $E_2$ \\ \hline
	 $1$   & $\underline{2.5}2218475780746$ & $\underline{3.57}619131917582$ & $\underline{4.68}971463368055$  \\
	 $3$   & $\underline{2.515}26712500729$ & $\underline{3.574}30507351366$ & $\underline{4.68897}621326843$  \\
	 $5$   & $\underline{2.51517}625626294$ & $\underline{3.5742971}5573030$ & $\underline{4.68897500}437660$  \\
	 $7$   & $\underline{2.515177}14074798$ & $\underline{3.574297136}92399$ & $\underline{4.68897500244}591$  \\ \hline
	TBA   & $\underline{2.51517709658632}$ & $\underline{3.57429713682553}$ & $\underline{4.68897500244662}$ \\ \hline 
	Numerical value & $2.51517709658632$ & $3.57429713682553$ & $4.68897500244662$ \\ \hline  
\end{tabular}
\caption{The first three eigenvalues of the modified Mathieu equation for $\hbar=1/2$. %The instanton number means the miximal value $n$ when truncating the sum \eqref{eq:F-NS}.
}  
\label{tab:spectrum1}
\end{center}
\end{table}

\begin{table}[tb]
\begin{center}
  \begin{tabular}{cccc}\hline
	Instanton number & $E_0$ & $E_1$ & $E_2$ \\ \hline
	 $1$   & $\underline{3.0}607381543889871$ & $\underline{5.285}4554447955830$ & $\underline{7.714}6870404010845$  \\
	 $3$   & $\underline{3.05917}25919787237$ & $\underline{5.28512}60103262340$ & $\underline{7.7145795}817132912$  \\
	 $5$   & $\underline{3.0591745}826455158$ & $\underline{5.285125967}0908562$ & $\underline{7.71457957299}08965$  \\
	 $7$   & $\underline{3.059174596}8723330$ & $\underline{5.2851259671795}903$ & $\underline{7.71457957299203}44$  \\ \hline
	TBA   & $\underline{3.0591745969015250}$ & $\underline{5.2851259671795203}$ & $\underline{7.7145795729920337}$ \\ \hline 
	Numerical value & $3.0591745969015250$ & $5.2851259671795203$ & $7.7145795729920337$ \\ \hline  
  \end{tabular}
\caption{The first three eigenvalues of the modified Mathieu equation for $\hbar=1$.}
 \label{tab:spectrum2}
\end{center}
\end{table}

Note that, in this four-dimensional case, the resummation of the WKB expansion provided by the NS free energy solves the spectral problem of the standard Toda lattice in terms of 
{\it convergent} series. As pointed out in \cite{kpt}, if we first expand in $\hbar$, as in (\ref{hbar-ex}), we recover the standard, asymptotic WKB expansion. In order to handle 
the quantization condition in this context, one needs to take into account possible non-perturbative effects, trans-series and their Borel resummation. However, the four-dimensional NS free energy  
provides a powerful treatment of the spectrum in terms of convergent series, and the machinery of resurgent trans-series is not really needed\footnote{The case of the Mathieu equation appears to be more complicated, see the discussion in \cite{kpt,krefl2,bd}.}. 

As we have tried to emphasize in this paper, the 5d story is different, due to the poles in the NS free energy. Starting from the all-orders WKB result, one can definitely follow the route of trans-series. 
However, in order to find a convergent series for the exact quantization condition, as in the 4d case, the resummation provided by the instanton partition function is not enough, and one 
needs the explicit non-perturbative contributions written down in (\ref{EQC}). 

\sectiono{Conclusions and outlook}

In this paper we have proposed exact 
quantization conditions for the relativistic Toda lattice. This has been a long-standing open problem, since solving the corresponding quantum Baxter equation has proved to be difficult. 
The insight of \cite{ns} is that one can use instead results from supersymmetric gauge theories and topological string theory. Inspired by the recent progress on non-perturbative aspects of quantum spectral curves \cite{km,ghm,wzh, hatsuda}, we have proposed a manifestly well-defined answer involving a convergent series. Our proposal reproduces 
the spectrum computed with numerical methods, at least in the cases $N=2$ and $N=3$. 

Clearly, it would be important to provide more evidence for our conjecture, and eventually to find a more rigorous justification for the simple ansatz (\ref{EQC}) 
presented in this paper. Physically, the non-perturbative corrections that we are adding seem to be due to complex instantons, as in the simpler setting of \cite{km}. It would be interesting 
to compute them from first principles in order to understand their origin and their S-duality structure in some detail.

There are clearly many avenues open for further research, since the relativistic Toda lattice admits many generalizations. One obvious extension of our conjecture
should be the Ruijsenaars-Schneider integrable system, 
which can be also engineered in topological string theory \cite{hiv}. As explicitly shown in \cite{efs} (see also \cite{marshakov,fm}), the relativistic Toda lattice is a particular 
example of the Goncharov--Kenyon construction, which associates a quantum integrable system to any 
toric CY manifold. It is natural to conjecture that (\ref{EQC}) provides an explicit, exact quantization condition for the Goncharov--Kenyon integrable system 
(in this more general setting, the matrix $C$ appearing in these equations should be replaced by the intersection matrix 
of the CY geometry considered in e.g. \cite{klemm-g2, cgm})\footnote{This conjecture has been 
presented and verified in detail in the subsequent work \cite{fhm}.}. Another interesting problem is the relation between the $g$ quantization conditions obtained for the relativistic Toda lattice (and also, presumably, for the Goncharov--Kenyon integrable system) and the 
approach based on quantizing the mirror curve presented in \cite{cgm}, which involves a single quantization condition encoded in the vanishing of a (quantum) theta function. 

We hope to address some of these problems in the near future.

\section*{Acknowledgements}
We would like to thank Andrea Brini, Santiago Codesido, Daniele Dorigoni, Vladimir Fock, 
Sebasti\'an Franco, Alba Grassi, Rinat Kashaev, Amir Kashani--Poor, Albrecht Klemm, Kazumi Okuyama, 
Samson Shatashvili, Junji Suzuki, J\"org Teschner and Szabolcs Zakany for useful 
discussions and correspondence. We are particularly thankful to Alba Grassi, Amir Kashani--Poor, 
Albrecht Klemm, Samson Shatashvili and J\"org Teschner for their 
comments on a preliminary version of this paper. This work is supported in part by the Fonds National Suisse, 
subsidies 200021-156995 and 200020-159581, and by the NCCR 51NF40-141869 ``The Mathematics of Physics" (SwissMAP). 

\appendix

%\sectiono{Some useful integrals}

\section{WKB analysis}\label{sec:WKB}
In this appendix, we review how to compute the quantum corrections to the classical periods \eqref{rt-actions}
by using the Baxter equation \eqref{q-baxter} \cite{gp,mir-mor}.
We consider the WKB ansatz of the wave function $\psi(\mu)$,
\be
\ba
\psi(\mu)&=\exp \left[ \frac{\ri}{\hbar} \int^\mu \rd \mu'\, P(\mu';\hbar_{\rm RT}) \right], \\
P(\mu;\hbar_{\rm RT})&=\sum_{n=0}^\infty \hbar_{\rm RT}^n P_n(\mu).
\ea
\ee
The Baxter equation \eqref{q-baxter} is of the same form as the difference equation studied in \cite{km}.
Plugging the WKB ansatz into the Baxter equation,
we find the leading order solution
\be
P_0(\mu)=2\cosh^{-1} \left| \frac{t(\re^{R\mu/2};R)}{R^N} \right|.
\ee
The quantum corrections can be fixed order by order.
Up to $n=2$, we find
\be
\ba
P_1(\mu)&=\left[ \frac{\ri}{2} \log \sinh P_0 \right]', \\
P_2(\mu)&=\frac{3}{8} \frac{\coth P_0}{\sinh^2 P_0} (P_0')^2-\frac{1}{12} \left( 1+\frac{3}{\sinh^2 P_0} \right)P_0''.
\ea
\ee
The quantum corrected action variables are then given by the period integrals
\be
I_k^\text{WKB}=\int_{{\cal I}_k} \rd \mu \, P(\mu; \hbar_\text{RT}).
\label{eq:I-WKB}
\ee
It turns out that the odd order parts $P_{2m+1}(\mu)$ are always written as a total derivative w.r.t. $\mu$,
and they do not contribute to the periods \eqref{eq:I-WKB}.
Therefore the quantum action variables have the following WKB expansion,
\be
I_k^\text{WKB}=\sum_{n=0}^\infty \hbar_\text{RT}^{2n} I_k^{(2n)}.
\ee
Of course, we have $I_k^{(0)}=I_k$ in \eqref{rt-actions}.
In this way, one can compute the quantum corrections $I_k^{(2n)}$ systematically.
However the integrands of the period integrals become very complicated, and it is hard to
see their analytic properties.
As explained in the main text, the large $H_i$ expansion (or small $z_i$ expansion) can be resummed
in all orders of $\hbar$. Each coefficient can be computed by the NS free energy.

\end{document}